\documentclass[12pt]{article} 
\usepackage{amsmath}
\usepackage{graphicx}
\usepackage{enumerate}
\usepackage{natbib}
\usepackage{multirow}
\usepackage{longtable}
\usepackage{booktabs}
\usepackage{amssymb}
\usepackage{bm}
\usepackage{mathrsfs}
\usepackage{amsthm}
\usepackage{amsfonts}
\usepackage{subfigure}
\usepackage{floatrow}
\usepackage{epsfig,amssymb,latexsym,verbatim}
\usepackage{epstopdf}
\usepackage{graphics}
\usepackage[english]{babel}
\usepackage[figuresright]{rotating}
\usepackage[dvipsnames]{xcolor}
\usepackage{color}
\usepackage{hyperref}

\usepackage{xr}

\allowdisplaybreaks
\newtheorem{theorem}{Theorem}

\newtheorem{remark}{Remark}

\newtheorem{corollary}{Corollary}
\numberwithin{equation}{section}
\numberwithin{lemma}{section}
\numberwithin{theorem}{section}
\numberwithin{prop}{section}
\numberwithin{corollary}{section}

\def\P{\mathbb{P}}
\def\E{\mathbb{E}}
\def\Var{\mbox{Var}}

\def\o{{\scriptstyle{\mathcal{O}}}}
\def\O{\mathcal{O}}

\newcommand{\blind}{0}

\begin{document}

\def\spacingset#1{\renewcommand{\baselinestretch}%
{#1}\small\normalsize} \spacingset{1}

\date{}
\if0\blind
{
  \title{\bf Unified theory of testing relevant hypotheses in
functional time series}
  \author{Leheng Cai\footnotemark[1] $\,$ and$\,$  Qirui Hu\footnotemark[2] \footnotemark[3] \hspace{.2cm}}
  \maketitle
  \renewcommand{\thefootnote}{\fnsymbol{footnote}}

  \footnotetext[1]{Department of Statistics and Data Science, Tsinghua University}
   \footnotetext[2]{School of Statistics and Data Science, Shanghai University of Finance and Economics}        
		\footnotetext[3]{Corresponding author: huqirui@mail.shufe.edu.cn}
} 

\if1\blind
{
  \bigskip
  \bigskip
  \bigskip
  \begin{center}
    { \bf Unified theory of testing relevant hypothesis in functional time series}
\end{center}
  \medskip
} \fi

\bigskip

\textbf{Abstract:} In this paper, we develop a {\em unified} framework for testing relevant hypotheses in functional time series. The proposed approach accommodates one-sample, two-sample, and change point problems for contaminated observations under arbitrary sampling schemes. Combining B-spline estimation with self-normalization, we construct nuisance-parameter-free tests that bypass auxiliary estimation of long-run covariance functions and measurement-error variance functions. We establish asymptotic validity by exploiting a sequential Gaussian approximation for dependent random vectors of moderately high dimension, which leads to a pivotal limiting distribution. We also provide sufficient conditions for the non-degeneracy of the self-normalizer and establish consistent decision rules. A key theoretical finding is that the proposed tests detect \(n^{-1/2}\)-local alternatives under arbitrary sampling frequencies. This uncovers a sparse-to-dense phase transition distinct from those typically observed in functional data analysis: while the sampling frequency affects the asymptotic variance, the detection rate remains \(n^{-1/2}\), even in sparsely sampled regimes. We further study multiple change  point alternatives and extend the theory to settings where consistent change point estimates are available. We also discuss the choice of self-normalizers, including the recently developed range-adjusted self-normalizer. Extensive simulations support the theoretical results, and applications to the AU.SHF implied volatility and traffic volume datasets demonstrate the practical utility of the proposed methods.

\textbf{Keywords:  Change point, Functional time series, Relevant hypotheses, Self-normalization, Unified inference.}

\maketitle
	
\newpage
\spacingset{1.9} 

\section{Introduction}

Functional time series analysis has emerged as a central topic in modern statistics, underpinning applications in neuroscience, economics, environmental science, and sociology; for standard lectures see \citet{bosq2000linear,horvath2012inference}.

\par In practice, trajectories are discretely observed with measurement error. Let the observed data be $\{(X_{ij},Y_{ij})\}_{i,j=1}^{n,N_i} $, where
\begin{align}\label{model1}
Y_{ij}= m(X_{ij}) + \xi_i(X_{ij}) + \sigma(X_{ij})\,\varepsilon_{ij}, \qquad i=1,\ldots,n,\; j=1,\ldots,N_i .
\end{align}
Here, $m(\cdot)$ denotes the mean function on $[0,1]$. The processes $\{\xi_i(\cdot)\}_{i=1}^{n}$ form a stationary sequence of square-integrable, mean-zero stochastic processes on $[0,1]$; equivalently, $\xi_i(\cdot)\in \mathcal L^2[0,1]$ almost surely,  $\mathbb{E}\!\int_0^1 \xi_i^2(x)\,dx<\infty$, and let $C_0(x,x'):=\E\xi_i(x)\xi_i(x')$ denote the covariance function. The design points $\{X_{ij}\}_{i,j=1}^{n,N_i}$ constitute the (possibly irregular) observation grid at which the random trajectories $\{\xi_i(\cdot)\}_{i=1}^{n}$ are recorded and are i.i.d. copies of a generic $X$ with density $f(\cdot)$. The measurement errors $\{\varepsilon_{ij}\}_{i,j=1}^{n,N_i}$ are i.i.d. with $\E(\varepsilon_{ij})=0$ and $\Var(\varepsilon_{ij})=1$, and the noise level varies with location through the variance function $\sigma^2(\cdot)$. Besides, \( \{N_{in}\}_{i=1}^{n} \) is an array of i.i.d. random variables,   following the common distribution \( N_{n} \), taking values in \( \mathbb{Z}_+ \), with \( \mathbb{P}(N_{in} > 1) > 0 \). For simplicity, we suppress the subscript \( n \) generally, that is, we write \( N_{in} \) as \( N_i \) and \( N_n \) as \( N \).
  Note that we do not require $\mathbb{E}(N_i)$ to be bounded, and it can grow to infinity  as $n\to \infty$. The sets $\{\xi_{i}(\cdot)\}_{i=1 }^{n }$, $\{X_{ij}\}_{i=1,j=1}^{n,N_i}$, and $\{\varepsilon_{ij}\}_{i=1,j=1}^{n,N_i}$  are mutually independent.

\par Early work in functional data focused on estimation \citep{yao2005functional,tonycai2011,ZW16,guo2025sparse}.  
Subsequent developments shifted attention toward inference, including dense
designs \citep{cao2012simultaneous,cai2024global},  sparse designs \citep{zheng2014smooth}, and  settings that bridge sparse and dense regimes \citep{cai2024sparse,cai2024detecting}.

\par From a statistical inference  standpoint, many inference problems in functional time series analysis can be cast as the hypothesis test
\begin{equation*}
H_0: \delta = 0 \quad \text{vs.} \quad H_1: \delta \neq 0,
\end{equation*}
where $\delta$ denotes an appropriate norm of a functional parameter, e.g., an $\mathcal{L}^2$- or $\mathcal{L}^{\infty}$-norm, which is formed from first- or second-order features of a stationary functional time series (such as the mean, covariance, or related components). For example, in the single change point problem, $\delta$ may denote the norm of the jump-magnitude function; under $H_0:\delta=0$ there is no change point.

\par However, as noted by \citet{BD87}, testing against arbitrarily small departures from zero need not be scientifically meaningful. This motivates tests for ``relevant'' differences,
\begin{equation}\label{hy2}
H_0: \delta \le \Delta \quad \text{vs.} \quad H_1: \delta > \Delta,
\end{equation}
where $\Delta>0$ is a pre-specified constant that encodes what constitutes a negligible deviation. The choice of norm and threshold $\Delta$ determines what is regarded as a meaningful effect and mitigates the classical consistency issue highlighted in \citet{B38}, whereby any nonzero effect is detected as $n\to\infty$. In change point problems, this focuses detection on substantively large jumps.
 By reversing  \eqref{hy2}, one obtains an ``equivalence'' formulation,
\begin{equation*}
H_0: \delta > \Delta \quad \text{vs.} \quad H_1: \delta \le \Delta,
\end{equation*}
which is used to assert similarity up to constant $\Delta$.

\par Relevant hypothesis testing has been explored widely, for example, in time series \citet{dette2016detecting}, and, more recently, for functional data; see \citet{dette2020functional} and \citet{dette2020testing}, which focus on the mean function of Banach-valued and Hilbert-valued trajectories, respectively, with various related topics; see \citet{dette2021bio}, \citet{van2022pivotal}, \citet{bucher2021deviations}, and \citet{van2024general}. 
These papers consider the relevant inference problem for fully observed trajectories, but it is not difficult to extend them to densely observed data via some nonparametric methods and proving the oracle efficiency. 

However, the central goal of this paper is to develop inference procedures for relevant hypotheses under arbitrary sampling schemes, ranging from sparse to dense. Existing methodologies are inadequate for this setting. In particular, individual observed or smoothed trajectories cannot be directly employed to construct moment-based estimators; rather, the observations must typically be pooled and analyzed through nonparametric methods. While the convergence rate of such pooled estimators generally depends on the relative order of the sampling frequency and the sample size \citep{ZW16,cai2024sparse}, it remains largely unclear whether a similar phenomenon persists for estimators of the target scalar functional. Moreover, nuisance functional parameters, such as the long-run covariance function, entail considerably more involved derivations and computations. This work successfully overcomes these difficulties by developing a sequential Gaussian approximation and employing a self-normalization approach. Although our work and \citet{dette2020testing} address the same fundamental statistical problem, our setting considers discretely observed data and incorporates nonparametric methodology, which introduces significant additional theoretical challenges for relevant inference. Finally, our framework does not require practitioners to pre-identify the underlying sampling regime in real data applications, thereby enabling relevant inference to be conducted seamlessly across sparse to dense regimes, ranging from idealized to fully realistic settings.

\par Our {\em unified} framework can be summarized in three parts:\\
• The framework includes one-sample, two-sample, and relevant change point problems, with further discussion of multiple change points. 
\\
• The framework is applicable to  arbitrary sampling schemes, and the resulting tests for relevant hypotheses can detect \(n^{-1/2}\)-local alternatives for functional time series observed under sampling regimes ranging from sparse to dense. By further analyzing the asymptotic variance of the estimators of the target scalar functionals, we identify the phase-transition boundary separating the sparse and dense regimes.\\
• The framework accommodates general dependence structures, including common linear and nonlinear stationary functional time series. We 
apply the recently developed sequential Gaussian approximation techniques and construct pivotal statistics via self-normalization, with further discussion of other  self-normalizers.
\par Surprisingly, although there are nontrivial differences between sparse and dense scenarios and between one-sample and relevant change point settings, our constructed pivotal statistics ultimately share the same fixed,  distribution-free asymptotic behavior, referred to as a {\em unified theory} for relevant hypothesis testing problems.

At the end of the introduction, we give a brief review of functional data analysis from sparse to dense, which is highly related to phase transitions. From an estimation perspective, the research begins with \citet{tonycai2011}, and \citet{ZW16} provides a state-of-the-art analysis for the mean and covariance functions under $\mathcal{L}^2$ and $\mathcal{L}^\infty$ norms, introducing the boundaries among sparse, semi-dense, and dense regimes, based on the attainment of   the $n^{-1/2}$ rate and the neglect of asymptotic bias. Subsequent work further explores the estimation of different components, such as \citet{berger2025dense}, \citet{sharghi2021mean}, \citet{Zhang2021UnifiedPC},  and \citet{guo2025sparse} for mean, mean derivatives, functional components, and non-asymptotic analysis of the mean and covariance functions, respectively.

From an inference perspective, one usually chooses an over-smoothed order in nonparametric methods to ignore bias and then conducts various inference tasks. The pioneering work starts with local inference for pointwise confidence intervals in \citet{kim2013unified}. Recently, global inference, such as the construction of simultaneous confidence bands, has been fully addressed;  see \citet{LWSTD2025} for kernel estimators, \citet{cai2024sparse} for series estimators, and \citet{cai2024detecting} for change point testing. In contrast to estimation problems, the phase-transition analyses in these papers emphasize the asymptotic behavior of the limiting/approximating processes rather than focusing merely on convergence rates; for our method, phase transitions are obtained by studying the asymptotic variance of the estimators. 
To the best of our knowledge, this paper is the first to address the relevant inference problem for functional time series from sparse to dense and to develop a unified framework.

\par The remainder of the paper is organized as follows: Section 2 establishes the unified methodology and asymptotic results for the one-sample and two-sample settings, and discusses self-normalization from sparse to dense regimes. Relevant change point problems, along with further discussion of multiple change point cases, are presented in Section 3. Extensive numerical experiments are given in Section 4. Section 5 illustrates the proposed method on real datasets. Some figures in Section 4, further discussions, additional simulations, all proofs and technical lemmas are provided in the supplementary materials.

\section{Relevant hypotheses in functional means}\label{sec:Relevant hypothesis in functional means}

For a real number $\nu\in(0,1]$ and integer $q\in\mathbb{N}$, $d\in\mathbb{N}_+$, write $\mathcal{H}_{q,\nu} [0,1]^d $ as the space of $(q,\nu)$-H\"{o}lder continuous functions on $[0,1]^d$, that is,
	\begin{align*}
		\mathcal{H}_{q,\nu} [0,1]^d =\left\{h:[0,1]^d\to \mathbb{R},~\left\|h\right\|_{q,\nu}=\sup_{t\neq s\in[0,1]^d ,\bm\alpha\in\mathbb N^d,\left\|\bm\alpha\right\|_1=q}\frac{\left|\partial^{\bm \alpha} h(t)-\partial^{\bm \alpha} h(s)\right|}{\left\|t-s\right\|_2^{\nu}}<\infty  \right\}.
	\end{align*}
To describe the temporal dependence, following \citet{zhou2023statistical} we specify $\xi_i(\cdot) = H(\cdot,\mathcal F_i)$, where $H: [0,1] \times \mathcal S^{\mathbb N} \to \mathbb R$ is a filter, $\mathcal F_i = (\cdots, e_{i-1}, e_i)$, and $\{e_i\}_{i \in \mathbb Z}$ is a sequence of i.i.d. random elements within some measurable space $\mathcal S$.  Then, the physical dependence measure   of $\{\xi_i(\cdot)\}_{i \in \mathbb Z}$ is defined by $ 
    \delta_H(i,r) = \sup_{x \in [0,1]}\left\|H(x, \mathcal F_i) - H(x, \mathcal F^\ast_i)\right\|_r$, 
where $\mathcal F^\ast_i = (\cdots, e_{-1}, e_0^\ast, e_1, \cdots, e_i)$, and $e_0^\ast$ is a copy of $e_0$ and independent of $\{e_i\}_{i \in \mathbb Z}$. The long-run covariance function of $\{\xi_i(\cdot)\}_{i=1}^{n}$ is expressed as $G(x,x^\prime) = \sum_{h=-\infty}^{\infty} \mbox{Cov}\left(H(x, \mathcal F_h), H(x^\prime, \mathcal F_0)\right).$
For two sequences $\{a_n\}$ and $\{b_n\}$, we write   $a_n\gtrsim b_n$ and $a_n\ll b_n$ to mean
that there exists  $c>0$  such that $a_n\geq  cb_n$ for all $n$, and $a_n=\o(b_n)$
respectively.

\subsection{One-sample problem}
 In the classical one-sample mean test, the objective is to assess whether the mean function $m(\cdot)$ is exactly equal to some pre-specified function $m_0(\cdot)$. Based on the discretely sampled observations $\{(X_{ij}, Y_{ij}) \}_{i,j=1}^{n,N_i}$,  our interest lies in the following testing problem under the framework of relevant hypotheses: \begin{align*}
     H_0:\left\|m(\cdot)-m_0(\cdot)\right\|_{\mathcal{L}^2}^2\leq \Delta,\quad H_1: \left\|m(\cdot)-m_0(\cdot)\right\|_{\mathcal{L}^2}^2>\Delta.
 \end{align*}
Without loss of generality, we consider the case where $m_0(\cdot) \equiv 0$, which leads to \begin{align}\label{one-sample-test}
    H_0:\left\|m(\cdot)\right\|_{\mathcal{L}^2}^2\leq \Delta,\quad H_1: \left\|m(\cdot)\right\|_{\mathcal{L}^2}^2>\Delta.
\end{align}
Otherwise, one  considers the data 
$\{(X_{ij}, \widetilde Y_{ij}) \}_{i,j=1}^{n,N_i}$ with 
$\widetilde{Y}_{ij} = Y_{ij} - m_0(X_{ij})$. 

Prior to elaborating on our methodology, B-spline functions are briefly described below. Let $\{t_l\}_{l=0}^{J_n+1}$ be a sequence of equally-spaced points, where $t_l = l / (J_n+1)$ for $0 \leq l \leq J_n+1$, which divide $[0,1]$ into $J_n+1$ equal sub-intervals, denoted as $I_l = [t_l, t_{l+1})$ for $l = 0, \ldots, J_n-1$, and $I_{J_n} = [t_{J_n}, 1]$. Let $\mathcal{S}^{p}_{J_n} = \mathcal{S}^{p}_{J_n}[0,1]$ be the polynomial spline space of order $p$ over $
\left\{I_l\right\}_{l=0}^{J_n}$, consisting of all functions that are $(p-2)$ times continuously differentiable on $[0,1]$ and are polynomials of degree $(p-1)$ within the sub-intervals $I_l$, $l = 0,\ldots,J_n$. Let $\bm B(x)=\{B_{1,p}(x),\ldots,B_{J_n+p,p}(x)\}^\top$ be the $p$-th order B-spline basis functions of $\mathcal{S}_{J_n}^{p}$, then $\mathcal{S}_{J_n}^{p}=\left\{
\sum_{l=1}^{J_n+p}\lambda_{l}B_{l,p}(x): \lambda_{l}\in \mathbb{R} \right\}$. Here, the dimension of $\bm B(x)$ depends on the number $J_n$ of interior  knots  and the order $p$, though this dependence is not reflected in the notation.  

A natural test statistic for (\ref{one-sample-test}) is $T_n =\left\|\widehat m(\cdot)\right\|_{\mathcal{L}^2}^2$, where $\widehat m(\cdot)$ is some nonparametric estimator of $m(\cdot)$. For  the  B-spline estimator \begin{align}\label{DEF:m_hat}
    \widehat{m}(\cdot)=\mathop{\arg\min}
\limits_{g(\cdot)\in \mathcal{S}_{J_n}^{p}}
\frac{1}{n}\sum_{i=1}^n \frac{1}{N_i}\sum_{j=1}^{N_i}  \left\{Y_{ij}-g
(X_{ij})\right\}^2,
\end{align}
one technically derives that   under some regular  conditions,  \begin{align*}
    \sqrt{n}/\{2 \tau_n (m)\}\left\{T_n-\left\|m(\cdot)\right\|_{\mathcal{L}^2}^2\right\}\xrightarrow{d} N\left(0,1\right),
\end{align*}
where $\tau_n^2(m) = \iint \bm B^\top(x)\mathbf\Sigma\bm B(x^\prime)m(x)m(x^\prime)dxdx^\prime$, and \begin{align*}
    \mathbf\Sigma &=  
    \mathbf{V}^{-1}  
    \E \left\{\bm B(X)\bm B^\top(X^\prime) G(X,X^\prime)\right\} \mathbf{V}^{-1} \\&+ \E (N^{-1})\mathbf{V}^{-1}\E \left[\bm B(X)\bm B^\top(X)\left\{C_0(X,X)+\sigma^2(X)\right\}  - \bm B(X)\bm B^\top(X^\prime) C_0(X,X^\prime)\right]\mathbf{V}^{-1}.
\end{align*}
Here, $X$ and $X^\prime$ are i.i.d. copies of $X_{ij}$, and $\mathbf V=\E\left\{ \bm B(X)\bm B^\top(X)\right\}$ is the inner product matrix.

Note that estimating the above asymptotic variance presents a great challenge, 
as it involves the long-run covariance function $G(\cdot,\cdot)$, the covariance function $C_0(\cdot,\cdot)$,  as well as the variance function $\sigma^2(\cdot)$ of the noise. Alternatively, we adopt the self-normalization (SN) approach to circumvent complicated nuisance-parameter estimation. The SN method was first proposed in \citet{shao2010self}; see the review \citet{shao2015self} for more details.
Given any $\epsilon\in(0,1)$ and for any $t\in[\epsilon,1]$, define the partial mean function by 
\begin{align*}
    \widehat{m}(t,\cdot) = \mathop{\arg\min}
\limits_{g(\cdot)\in \mathcal{S}_{J_n}^{p}}
\frac{1}{[nt]}\sum_{i=1}^{[nt]} \frac{1}{N_i}\sum_{j=1}^{N_i}  \left\{Y_{ij}-g
(X_{ij})\right\}^2.
\end{align*}
A commonly used self-normalizer  is introduced as follows: \begin{align*}
V_{n,\epsilon}=\left[\int_\epsilon^1 t^4\left\{  \int_0^1 \widehat m^2(t,x)dx-\int_0^1   \widehat m^2(x)dx\right\}^2dt\right]^{1/2}. 
\end{align*}
Then, the hypothesis in (\ref{one-sample-test}) can be rejected if the following  holds: \begin{align}\label{one-sample-rule}
    (T_n-\Delta)/V_{n,\epsilon}> Q_{1-\alpha,\epsilon},
\end{align}
in which $Q_{1-\alpha,\epsilon}$ is the upper-$\alpha$ quantile of the pivotal distribution  $
   \mathcal W(1)/\left[ \int_\epsilon^1 t^2\{\mathcal W(t)-t\mathcal W(1)\}^2 dt \right]^{1/2},
$
where $\mathcal W(\cdot)$ is a standard Brownian motion. 

The intuition behind the above testing procedure relies on the following joint weak convergence under some regular conditions:  
\begin{align}\label{eq:rau_n(m)}
    \left(\frac{\sqrt{n}\left\{T_n-\left\|m(\cdot)\right\|_{\mathcal{L}^2}^2\right\}}{2 \tau_n (m)},\frac{\sqrt{n}V_{n,\epsilon}}{2 \tau_n (m)}  \right)\xrightarrow{d} \left( \mathcal W(1),\left[ \int_\epsilon^1 t^2\{\mathcal W(t)-t\mathcal W(1)\}^2 dt \right]^{1/2}\right).
\end{align}
According to the continuous mapping theorem, taking the ratio of the two statistics on the left hand side effectively cancels out the nuisance scaling factor $\tau_n (m)$. 

To rigorously investigate asymptotic properties of the proposed testing procedure, some mild assumptions are imposed.  
\begin{itemize}
    \item[(A1)] For some integer $q>0$ and some positive real number $\nu\in(0,1]$, $m(\cdot)\in \mathcal{H}_{q,\nu}[0,1]$. In the following, denote  $q^\ast=q+\nu$.
		\item[(A2)] The density function $f(\cdot)$ of $X$ is bounded above and away from zero, i.e., $c_{f}\leq \inf_{x\in[0,1]} f(x)  \leq \sup_{x\in[0,1]} f(x) \leq C_{f}$ for some positive constants $c_{f}$ and $C_{f}$.
  \item[(A3)] 
  The variance function $\sigma^2(\cdot)$  is uniformly bounded, and  $\E|\varepsilon_{ij}|^{r_1}<\infty$ for some integer $r_1\geq 3$.
  \item[(A4)] The  function $G(\cdot,\cdot) \in\mathcal{H}_{0,\varpi}[0,1]^2$ for some $\varpi\in (0,1]$,   
 and $\sup_{x\in[0,1]}\E|\xi_i(x)|^{r_2}<\infty$ for some integer $r_2\geq 3$.
  \item[(A5)]
For some $\beta\geq 3$ and $r_2$ in (A4), the dependence measure $\delta_H(i,r_2)=\O\left(i^{-\beta}\right)$. 
\item[(A6)] Suppose that the spline order is $p\geq q^\ast$, and let $r=\min\{r_1,r_2\}$.  
\begin{align*}
    &J_n^{-q^\ast}n^{1/2}  =\o(1), \,\,\,  n^{-1/2}J_n   \log^{1/2}(n)=\o(1),\\& n^{-1/2}J_n^{3/2 }E(N^{-1}) \log^{1/2}(n)=\o(1),\\
    &n^{-\frac{r-2}{3r-2}} J_n^{\frac{7r-6}{3r-2 }}\left\{\sum_{s=1}^{r}\E \left(N^{s-r}\right) J_n^{-s}\right\}^{2/r}   \log(n)=\o(1).
\end{align*} 
 
\end{itemize}

The smoothness requirements on the mean functions and the long-run covariance function in Assumptions (A1) and (A4) are standard in the literature; see, for instance, \citet{2022lijie} and \citet{cai2024detecting}. Assumption (A2) is a common condition for random design settings, as in \citet{tonycai2011} and \citet{guo2025sparse}. Assumptions (A3) and (A4) specify moment conditions for the measurement error $\varepsilon_{ij}$ and the stochastic process $\xi_i(\cdot)$, with related conditions discussed in \citet{ZW16} and \citet{Zhang2021UnifiedPC}. 
Assumption (A5) prescribes a polynomial decay rate for the physical dependence measure, ensuring weak temporal dependence in the functional time series. Similar polynomial rates can be found in 
\citet{zhou2023statistical}. The parameter ranges in Assumptions (A1), (A3), and (A4) are detailed in Assumption (A6). For illustration, we consider the following simple setup, $q^\ast=q+\nu=4$,   $r_1=r_2\geq 6$, $N_i\equiv N\geq 2$ for all $i=1,2,\ldots,n$.  One may employ cubic splines with $p = 4$ and choose the number of knots $J_n$ such that
 $ n^{1/8} \ll J_n\ll \min\{n^{3/23}N^{20/23}(\log n)^{-12/23}, n^{1/3}(\log n)^{-1/3}\}$. 

To prevent the first-order asymptotic expansion of the normalizer $V_{n,\epsilon}$  from degenerating, we impose the following condition $(\ast)$: $\tau_n^2(m)\gtrsim 1$.

\begin{remark}\label{remark1}
For any   function $g\in\mathcal C[0,1]$, denote 
$$
    \tau^2(g) = \iint G(x,x^\prime)g(x)g(x^\prime)dxdx^\prime.
$$ 
The validity of $(\ast)$ can be ensured by the following sufficient condition: Under Assumptions (A2)-(A4), 
when $ \E(N^{-1})$ is bounded away from zero, assume that either $\tau^2(m) > 0$ or
$\inf_{x\in[0,1]} \sigma^2(x)  \geq c_{0}$ 
for some positive constant $c_{0}$; when $ \E(N^{-1})\to 0$, assume that  $\tau^2(m) > 0$. Note that   $\tau^2(m) > 0$ is guaranteed to hold provided that $m$ has a nonzero projection onto the eigenspace of $G$ associated with positive eigenvalues and is also imposed in \citet{dette2020testing}. 
Hence, Theorem 1 of \citet{dette2020testing} is a special case of Theorem \ref{THM:one-sample}. In particular, when $C_0(x,x)\equiv 0$, our framework reduces to nonparametric regression; when $\sigma^2(x)\equiv 0$, the trajectories are observed without measurement error, recovering the results of \citet{dette2020testing} as $\mathbb{E}(N^{-1})\to 0$.
\end{remark}

\par The following Theorem \ref{THM:one-sample}  provides  theoretical justification for   asymptotic properties of the decision rule in $(\ref{one-sample-rule})$. 
\begin{theorem}\label{THM:one-sample}
Suppose that $\Delta>0$. Under Assumptions (A1)-(A6), the  rejection rule in (\ref{one-sample-rule}) satisfies that 
\begin{align*}
\lim_{n\to\infty}\P\left(T_n>\Delta+Q_{1-\alpha,\epsilon}V_{n,\epsilon}\right)= \begin{cases}
0  & \text{ if }  \int m^2(x)dx< \Delta, \\
\alpha  & \text{ if } \int m^2(x)dx = \Delta \text{ and } (\ast),  \\
1  & \text{ if } \int m^2(x)dx> \Delta.
\end{cases} 
\end{align*}
\end{theorem}

By Theorem~\ref{THM:one-sample}, the SN-based testing rule in
\eqref{one-sample-rule} is asymptotically valid. At the boundary of the
null hypothesis, where \(\int m^2(x)\,dx=\Delta\), condition \((\ast)\) is
needed to guarantee that the asymptotic size of the test equals
\(\alpha\). The reason is that the nuisance scaling factor \(\tau_n(m)\),
which is removed through the self-normalized statistic in
\eqref{eq:rau_n(m)}, has to dominate the Gaussian approximation error in
Lemma~S.4.9. 
Without this dominance condition,
the denominator \(\sqrt{n}V_{n,\epsilon}\) may degenerate. In contrast,
once this condition is imposed, \(\sqrt{n}V_{n,\epsilon}\) remains bounded
in probability. This property further implies that the proposed test is
capable of detecting \(n^{-1/2}\)-local alternatives, as stated in
Theorem~\ref{THM:onesample_localH1}.

\begin{theorem}\label{THM:onesample_localH1}
    Suppose  that Assumptions (A1)-(A6)  and $(\ast)$ hold. For any significance level $\alpha\in(0,1)$, there exists some positive constant $c_0$ such that under   the local alternatives  $H_{1n}:\left\|m_n(\cdot)\right\|_{\mathcal{L}^2}^2=\Delta+c_0/\sqrt{ n},$  the decision rule (\ref{one-sample-rule}) has nontrivial power $
\liminf_{n\to\infty}\P\left(T_n>\Delta+Q_{1-\alpha,\epsilon}V_{n,\epsilon}\right)  >\alpha
$. 
\end{theorem}

The SN-based statistic \((T_n-\Delta)/V_{n,\epsilon}\) has a
distribution-free asymptotic law on the boundary between the null and
alternative hypotheses in \eqref{one-sample-test}, and the resulting test
has nontrivial asymptotic power against \(n^{-1/2}\)-local alternatives.
We therefore further examine the asymptotic variance of \(T_n\) to
quantify the impact of the sampling frequency
\(\{\E(N^{-1})\}^{-1}\), which reveals a sparse-to-dense
phase-transition phenomenon.

 \begin{theorem}\label{THM:taun}
    Suppose  that Assumptions (A1)-(A6) and $\{\E(N^{-1})\}^{-1}\to \infty$  as $n\to\infty$. Then, \begin{align*}
        \lim_{n\to\infty}\tau_n^2(m)=\tau^2(m).
    \end{align*}
\end{theorem}

 
\begin{corollary}\label{COR:onesample}
Suppose  that  Assumptions (A1)-(A6) and $(\ast)$ hold. \begin{itemize}
    \item[(1)] When $\{\E(N^{-1})\}^{-1}=\O(1)$ and  $\sigma^2(\cdot)\neq 0$,   the asymptotic variance of $T_n$ is strictly larger than $4\tau^2(m)/n$, referred to as ``sparse''.
    \item[(2)] When $\{\E(N^{-1})\}^{-1}\to\infty$,  the asymptotic variance of $T_n$ is equal to $4\tau^2(m)/n$, referred to as ``dense''.
\end{itemize}
\end{corollary}

Existing phase-transition results in the functional data literature
\citep{cai2024detecting,cai2024sparse} typically require the effective
sampling frequency \(\{\E(N^{-1})\}^{-1}\) to dominate the number of spline
knots \(J_n\) for the corresponding statistic to attain the \(n^{-1/2}\)
convergence rate. In contrast, for the relevant testing problem considered
here, Corollary~\ref{COR:onesample} shows that \(T_n\) attains the
\(n^{-1/2}\) rate under arbitrary sampling frequencies. Moreover, as long as
\(\{\E(N^{-1})\}^{-1}\to\infty\), neither the sampling frequency nor the
contamination induced by discrete observations enters the first-order
asymptotic behavior of \(T_n\).

It is important to emphasize that we do not claim that
\(\widehat m(\cdot)\) itself converges at the \(n^{-1/2}\) rate with respect
to the \(\mathcal L^2\)-norm, the \(\mathcal L^\infty\)-norm, or pointwise.
The parametric rate is obtained only for the integrated scalar functional
\(T_n\). Indeed, the leading term in the asymptotic expansion of
\(T_n-\|m(\cdot)\|^2\) is the integral inner product between
\(\widehat m(\cdot)-m(\cdot)\) and \(2m(\cdot)\). Intuitively, this
integration averages out local estimation errors, allowing positive and
negative fluctuations to cancel. Consequently, \(T_n\) may exhibit an
\(n^{-1/2}\) convergence rate even when \(\widehat m(\cdot)\) itself does
not. Related examples in which integral functionals admit parametric-rate
inference can be found in
\citep{bickel1988estimating,6dd3c0f4-00c9-3011-90a1-202033e64598,zhou2018efficient}.

\par  Additionally, we emphasize that  our methodology is not sensitive to the choice of $\epsilon$  and the self-normalizers. Besides, from a practical perspective, the nonparametric methods are not restricted to B-splines and can be further extended to orthogonal series regression. Due to the space limits, we put these discussions and additional numerical simulations in Sections S.1
and S.3
of the Supplementary Materials. 

\subsection{Extension to two-sample problems}
It is often of interest to compare mean functions of two stationary functional time series under the relevant hypothesis framework. 
For $d=1,2$,  let $\{X_{dij},Y_{dij}\}_{i,j=1}^{n_d,N_{di}}$ denote the discretely observed data from  two mutually independent populations,
\begin{align*}
    Y_{dij} = m_d(X_{dij})+\xi_{di}(X_{dij})+\sigma_d(X_{dij})\varepsilon_{dij},\,\,1\leq i\leq n_d,\,\, 1\leq j\leq N_{di},
\end{align*}
and we consider the following two-sample relevant hypothesis:  \begin{align}\label{two-sample-test}
    H_0:\left\|m_{1}(\cdot)-m_{2}(\cdot)\right\|_{\mathcal{L}^2}^2\leq \Delta,\quad H_1: \left\|m_{1}(\cdot)-m_{2}(\cdot)\right\|_{\mathcal{L}^2}^2>\Delta.
\end{align}
For $d=1,2$ and $t\in [\epsilon,1]$, define  
\begin{align*}
    \widehat{m}_d(t,\cdot) = \mathop{\arg\min}
\limits_{g_d(\cdot)\in \mathcal{S}_{J_{n_d}}^{p_d}}
\frac{1}{[n_d t]}\sum_{i=1}^{[n_d t]} \frac{1}{N_{di}}\sum_{j=1}^{N_{di}}  \left\{Y_{dij}-g_d
(X_{dij})\right\}^2,
\end{align*}
and let $\widehat m_d(\cdot)=\widehat m_d(1,\cdot)$. 
Further define $T_{n_1,n_2}^{\dagger} =\int_0^1 \left\{\widehat m_1( x)-\widehat m_2(x)\right\}^2dx$, and \begin{align*}
V_{n_1,n_2,\epsilon}^{\dagger}=\left(\int_\epsilon^1 t^4\left[  \int_0^1 \left\{\widehat m_1(t,x)-\widehat m_2(t,x)\right\}^2 dx-\int_0^1 \left\{  \widehat m_1(x)-\widehat m_2(x)\right\}^2dx\right]^2dt\right)^{1/2}.
\end{align*}
The hypothesis in (\ref{two-sample-test}) can be rejected if the following  holds: \begin{align}\label{two-sample-rule}
T_{n_1,n_2}^\dagger>\Delta+Q_{1-\alpha,\epsilon}V^\dagger_{n_1,n_2,\epsilon}.
\end{align}

For $d=1,2$, let $N_{(d)}$ denote an i.i.d. copy of $\{N_{di}\}_{i=1}^{n_d}$, $C_{d,0}(\cdot,\cdot)$ and $G_{d}(\cdot,\cdot)$ denote the covariance and the long-run covariance function of $\left\{\xi_{d i}(\cdot)\right\}_{i=1}^{n_d}$, respectively.  
Define   $\mathbf{V}_d = \E \bm B_d(X_d)\bm B_d^\top(X_d)$, and
\begin{equation*}
    \begin{aligned}
     \mathbf \Sigma_{d}^{\dagger}&=
    \mathbf{V}_d^{-1} \E \left\{\bm B_d(X_d)\bm B_d^\top(X_d^\prime) G_d(X_d,X_d^\prime) \right\} \mathbf{V}_d^{-1} \\&+
    \E (N_{(d)}^{-1}) \mathbf{V}_d^{-1} \E \left[\bm B_d(X_d)\bm B_d^\top(X_d)\left\{C_{d,0}(X_d,X_d)+\sigma_d^2(X_d)\right\} \right.\\&\left.- \bm B_d(X_d)\bm B_d^\top(X_d^\prime)C_{d,0}(X_d,X_d^\prime)\right]\mathbf{V}_d^{-1}.
    \end{aligned}
\end{equation*}
Further define  \begin{align*}
    \tau_{n_1,n_2}^{\dagger2}(g)=\iint \left\{ \frac{1}{n_1}\bm B_1^\top(x)\mathbf\Sigma_1^{\dagger}\bm B_1(x^\prime)+\frac{1}{n_2}\bm B_2^\top(x)\mathbf\Sigma_2^{\dagger}\bm B_2(x^\prime) \right\}g(x)g(x^\prime) dxdx^\prime.
\end{align*}
To mimic the condition $(\ast)$, the following condition $(\ast\ast)$ is considered: 
$$\tau_{n_1,n_2}^{\dagger2}(m_1-m_2)\gtrsim n_1^{-1}+n_2^{-1}.\qquad(\ast\ast)$$

Theorem  \ref{THM:twosample} below establishes the asymptotic correctness of the rule in (\ref{two-sample-rule}). 
\begin{theorem}\label{THM:twosample}
Suppose that $\Delta>0$. 
If Assumptions (A1)-(A6) are satisfied for data $\{X_{dij},Y_{dij}\}_{i,j=1}^{n_d,N_{di}}$  and the choices of smoothing parameters $J_{n_d}$, $d=1,2$,  respectively, the  rejection rule in (\ref{two-sample-rule}) satisfies that 
\begin{align*}
\lim_{n_1,n_2\to\infty}\P\left(T_{n_1,n_2}^{\dagger}>\Delta+Q_{1-\alpha,\epsilon}V_{n_1,n_2,\epsilon}^{\dagger}\right)= \begin{cases}
0  & \text{ if }  \left\|m_{1}(\cdot)-m_{2}(\cdot)\right\|_{\mathcal{L}^2}^2< \Delta, \\
\alpha  & \text{ if } \left\|m_{1}(\cdot)-m_{2}(\cdot)\right\|_{\mathcal{L}^2}^2 = \Delta \text{ and }    (\ast\ast),  \\
1  & \text{ if } \left\|m_{1}(\cdot)-m_{2}(\cdot)\right\|_{\mathcal{L}^2}^2> \Delta.
\end{cases} 
\end{align*}
\end{theorem}

\begin{remark}\label{remark2} Similar to Remark 1, the condition $(\ast\ast)$ can be implied by some high-level conditions.
  For any function $g\in\mathcal C[0,1]$, denote 
$$
    \tau^2_d(g) = \iint G_d(x,x^\prime)g(x)g(x^\prime)dxdx^\prime,
$$ 
where  $G_d(\cdot,\cdot)$ is the long-run covariance function of $\{\xi_{di}(\cdot)\}_{i\in\mathbb Z}$, and   let $d_0=\mathop{\arg\min}
\limits_{d=1,2} n_d  $. Then, a sufficient condition for $(\ast\ast)$ to hold is that: Under Assumptions (A2)-(A4) for the $d_{0}$-th sample, when $ \E(N_{(d_0)}^{-1}) $ is bounded away from zero, assume that either $\tau_{d_0}^2(m_1-m_2) > 0$ or 
$\inf_{x\in[0,1]}  \sigma_{d_0}^2(x) \geq c_{0}$ 
for some positive constant $c_{0}$;  
when $ \E(N_{(d_0)}^{-1})\to 0$, assume that  $\tau_{d_0}^2(m_1-m_2) > 0$. 

\end{remark}

It is  worth noting our setup allows unbalanced sample sizes and sampling frequencies between the two groups, and the smaller group determines the asymptotic behavior. 
Besides, the smoothing parameters $J_{n_1}$ and  $J_{n_2}$  for estimating the mean functions of the two samples may be different, provided that each is chosen according to the respective orders of $n_d$ and $\E(N_{di}^{-1})$ specified in Assumption (A6) with slight modifications to the notation.  
Mimicking the results in Theorem \ref{THM:onesample_localH1},  one can show that, in the two-sample setting,   the detection rate of the test is given by 
$
\max\left\{n_1^{-1/2},n_2^{-1/2}\right\} 
$ under arbitrary sampling schemes.

\section{Testing relevant change points}

\subsection{Single change point}
In this section, we slightly modify model (\ref{model1}) to accommodate a change point structure:
\begin{align*}
Y_{ij} = m_i(X_{ij}) + \xi_i(X_{ij}) + \sigma(X_{ij})\,\varepsilon_{ij}, 
\qquad i=1,\ldots,n,\; j=1,\ldots,N_i.
\end{align*}
Here, $\{m_i(\cdot)\}_{i=1}^{n}$ represents a sequence of mean functions defined on $[0,1]$. 
Assume that there is an abrupt change point in the mean functions $\{m_i(\cdot)\}_{i=1}^{n}$, i.e.,
$m_1(\cdot)=\cdots=m_{k_n}(\cdot)\neq   m_{k_n+1}(\cdot)=\cdots=m_{n}(\cdot),$ 
and jump magnitude function $\delta(\cdot) = m_{k_n+1}(\cdot)-m_{k_n}(\cdot)$. 
We aim  to address  the following relevant change point testing,   which is of great practical interest  as discussed in~\citet{DK21}:   
\begin{align}\label{relevantchange-test}
    H_0:\left\|\delta(\cdot) \right\|_{\mathcal{L}^2}^2\leq \Delta,\quad H_1: \left\|\delta(\cdot)\right\|_{\mathcal{L}^2}^2>\Delta.
\end{align}
Given a consistent estimator $\widehat{k}_n$ of the change point $k_n$, for $t \in [\epsilon,1]$  define
\begin{align*}
    &\widehat{\mu}_1(t,\cdot) = \mathop{\arg\min}
\limits_{g_1(\cdot)\in \mathcal{S}_{J_{n}}^{p}}
\frac{1}{[\widehat k_n t]}\sum_{i=1}^{[\widehat k_n t]} \frac{1}{N_{i}}\sum_{j=1}^{N_{i}}  \left\{Y_{ij}-g_1
(X_{ij})\right\}^2,\\&
\widehat{\mu}_2(t,\cdot) = \mathop{\arg\min}
\limits_{g_2(\cdot)\in \mathcal{S}_{J_{n}}^{p}}
\frac{1}{[(n-\widehat k_n) t]}\sum_{i= \widehat k_n  +1}^{\widehat k_n+ [ (n-\widehat k_n)t]} \frac{1}{N_{i}}\sum_{j=1}^{N_{i}}  \left\{Y_{ij}-g_2
(X_{ij})\right\}^2.
\end{align*}
Our suggested estimator $\widehat k_{n,\mathcal{L}^2}$  is a canonical choice:  
\begin{align*}
    \widehat k_{n,\mathcal{L}^2} = \mathop{\arg\max}
\limits_{\epsilon n\leq k\leq (1-\epsilon)n} \left\|  \bm B^\top(\cdot)\widehat{\mathbf Q}^{-1}\left\{ \sum_{i=1}^{k}\frac{1}{N_i}\sum_{j=1}^{N_i}\bm B(X_{ij})Y_{ij}- \frac{k}{n}\sum_{i=1}^{n}\frac{1}{N_i}\sum_{j=1}^{N_i}\bm B(X_{ij})Y_{ij}
 \right\}\right\|_{\mathcal{L}^2}^2,
\end{align*}
in which $
    \widehat{\mathbf Q} = n^{-1}\sum_{i=1}^{n}N_i^{-1}\sum_{j=1}^{N_i}\bm B(X_{ij})\bm B^\top(X_{ij}).
$
The consistency of $\widehat{k}_{n,\mathcal{L}^2}$ is established in \citet{cai2024detecting}, where it is shown to satisfy Assumption (B2) stated below under mild conditions. 
For notational simplicity, we denote  $\widehat\delta(t,x)=\widehat \mu_1(t,x)-\widehat \mu_2(t,x)$, and further define \begin{align*}
    S_n =\int_0^1  \widehat \delta^2(1,x) dx,\quad 
    U_{n,\epsilon}=\left[\int_\epsilon^1 t^4\left\{  \int_0^1 \widehat \delta^2(t,x)dx-\int_0^1   \widehat \delta^2(1,x)dx\right\}^2dt\right]^{1/2}.
\end{align*}
The hypothesis in (\ref{relevantchange-test}) can be rejected if the following  holds: \begin{align}\label{change-point-rule}
S_n>\Delta+Q_{1-\alpha,\epsilon}U_{n,\epsilon}.
\end{align}

\begin{itemize}
    \item[(B1)]   For some integer $q>0$ and some positive real number $\nu\in(0,1]$, $m_i(\cdot)\in \mathcal{H}_{q,\nu}[0,1]$, $i=k_n,k_n+1$.
    \item[(B2)] The change point estimator $\widehat k_n$ satisfies that $\left|\widehat k_n-k_n\right|=\o_p\left(n^{1/2} \right)$. 
\end{itemize}
{\color{black}In order to mirror condition $(*)$, we impose the following condition $(\star)$: $\tau_n^2(\delta) \gtrsim 1$, where $\tau_n^2(\delta) = \iint \bm B^\top(x)\mathbf\Sigma\bm B(x^\prime)\delta(x)\delta(x^\prime)dxdx^\prime$. A sufficient condition to ensure $(\star)$ holds is to replace $m(\cdot)$ in Remark \ref{remark1} with $\delta(\cdot)$.} 
Theorem \ref{THM:change-point} offers a rigorous theoretical foundation for the asymptotic validity of the decision rule in 
(\ref{change-point-rule}).

\begin{theorem}
    \label{THM:change-point}
    Suppose that $\Delta>0$ and $ 
\rho:=\lim_{n\to\infty} k_n/n \in(\epsilon,1-\epsilon)$. Under Assumptions (A2)-(A6) and (B1)-(B2), the  rejection rule in (\ref{change-point-rule}) satisfies that 
\begin{align*}
    \lim_{n\to\infty}\P\left(S_n>\Delta+Q_{1-\alpha,\epsilon}U_{n,\epsilon}\right)= \begin{cases}
0  & \text{ if }  \int \delta^2(x)dx< \Delta, \\
\alpha  & \text{ if } \int \delta^2(x)dx = \Delta \text{ and }   (\star),  \\
1  & \text{ if } \int \delta^2(x)dx> \Delta.
\end{cases} 
\end{align*}
\end{theorem}

The main idea of Theorem \ref{THM:change-point} is similar to that of Theorem \ref{THM:one-sample}; however, meticulous and technical analysis is more involved, as additional efforts are required to verify that the error introduced by the change point estimator is asymptotically negligible. 

The following Theorem \ref{THM:changepoint_localH1} and Corollary \ref{COR:changepoint} show that relevant change‐point testing shares the same detection rate as the one‐sample relevant testing, and possesses the same phase transition boundary conditions.  

\begin{theorem}\label{THM:changepoint_localH1}
    Suppose  that $(\star)$ and Assumptions (A2)-(A6) and (B1)-(B2) hold. For any significance level $\alpha\in(0,1)$, there exists some constant $c_0>0$ such that under   the local alternatives: $H_{1n}:\left\|\delta_n(\cdot)\right\|_{\mathcal{L}^2}^2=\Delta+c_0/\sqrt{n},$ the decision rule (\ref{change-point-rule}) has nontrivial power $
    \liminf_{n\to\infty}\P\left(S_n>\Delta+Q_{1-\alpha,\epsilon}U_{n,\epsilon}\right)  >\alpha$. 
\end{theorem}


\begin{corollary}\label{COR:changepoint}
Suppose  that $(\star)$ and Assumptions (A2)-(A6) and (B1)-(B2) hold. \begin{itemize}
\item[(1)] When $\{\E(N^{-1})\}^{-1}=\O(1)$ and  $\sigma^2(\cdot)\neq 0$,   the asymptotic variance of $S_n$ is strictly larger than $4\tau^2(\delta)/\{n\rho(1-\rho)\}$, referred to as ``sparse''.
    \item[(2)] When $\{\E(N^{-1})\}^{-1}\to\infty$,  the asymptotic variance of $S_n$ is equal to $4\tau^2(\delta)/\{n\rho(1-\rho)\}$, referred to as ``dense''.
\end{itemize}
\end{corollary}

\subsection{Extension to multiple change points}
\par 
The single relevant change point setting can be naturally extended to the multiple change point scenario. 
We now consider the case where there are $K$ change points, and let $0=\theta_0<\theta_1<\ldots<\theta_K<\theta_{K+1}=1$ denote the locations of unknown change points, and $\delta_k(\cdot)=m_{[n\theta_k]+1}(\cdot)-m_{[n\theta_k]}(\cdot)\neq 0$ for $1\leq k\leq K$. 

To tackle the following multiple relevant change point testing problem,  \begin{align}\label{H0:multiple}
    H_0:\sum_{k=1}^{K}\left\|\delta_k\right\|_{\mathcal{L}^2}^2\leq \Delta,\quad H_1:\sum_{k=1}^{K}\left\|\delta_k\right\|_{\mathcal{L}^2}^2> \Delta, 
\end{align}
we first require consistent estimators for the change point locations $\theta_k$, denoted by $\widehat{\theta}_k$. For instance, one may employ the methodology developed in~\citet{madrid2022change}, which provides both the estimation algorithm and theoretical guarantees. 
Then, 
we denote $\widehat n_k=\left[n\widehat\theta_k\right]-\left[n\widehat\theta_{k-1}\right]$ 
for $1\leq k\leq K+1$,  and for $t\in[\epsilon,1]$ define  
\begin{align}\label{mu_k}
    &\widehat{\mu}_k(t,\cdot) = \mathop{\arg\min}
\limits_{g_k(\cdot)\in \mathcal{S}_{J_{n}}^{p}}
\frac{1}{\left[t\widehat n_{k}\right]}\sum_{i=\left[n\widehat\theta_{k-1}\right]+1}^{\left[t\widehat n_{k}\right]+\left[n\widehat\theta_{k-1}\right]} \frac{1}{N_{i}}\sum_{j=1}^{N_{i}}  \left\{Y_{ij}-g_k
(X_{ij})\right\}^2. 
\end{align}
Further define $\widehat\delta_k(t,\cdot)=\widehat\mu_{k+1}(t,\cdot)-\widehat\mu_{k}(t,\cdot)$ for $1\leq k\leq K$. 
The hypothesis in (\ref{H0:multiple}) can be rejected if the following  holds: \begin{align}\label{multichange-point-rule}
S_n^{\dagger}>\Delta+Q_{1-\alpha,\epsilon}U_{n,\epsilon}^{\dagger},
\end{align}
where $S_n^{\dagger} =\sum_{k=1}^{K}\int_0^1  \widehat \delta_k^2(1,x) dx$, and \begin{align*}
U_{n,\epsilon}^{\dagger}=\left[\int_\epsilon^1 t^4\left\{  \sum_{k=1}^{K}\int_0^1 \widehat \delta_k^2(t,x)dx-\sum_{k=1}^{K}\int_0^1   \widehat \delta_k^2(1,x)dx\right\}^2dt\right]^{1/2}.
\end{align*}
The following conditions are adapted from Assumptions (B1)-(B2), with appropriate modifications to facilitate the theoretical justification. 
\begin{itemize}
    \item[(B1$^{\prime}$)]   For some integer $q>0$ and some positive real number $\nu\in(0,1]$, $m_i(\cdot)\in \mathcal{H}_{q,\nu}[0,1]$ for $i=1,\ldots,n$.
    \item[(B2$^{\prime}$)] The change point estimator $\widehat \theta_k$ satisfies that $\left|\widehat \theta_k-\theta_k\right|=\o_p\left(n^{-1/2} \right)$.  
\end{itemize}

{\color{black} Assumption (B2$^{\prime}$) imposes a requirement on the convergence rate of the change point estimators $\widehat \theta_k$'s, a condition that is guaranteed by Theorem 1 of \citet{madrid2022change}.}
{\color{black}To parallel condition $(\star)$, we impose the following requirement $(\star\star)$:  
$\sum_{k=1}^{K}\tau_n^2(\delta_k) \gtrsim 1 $. 
It suffices that there exists some $1 \leq k \leq K$ such that condition $(\star)$ holds when $\delta(\cdot)$ is replaced by $\delta_k(\cdot)$.
}
Then, one  extends Theorem \ref{THM:change-point}  to the following multiple relevant change points version. 
\begin{theorem}
    \label{THM:multi-change-point}
    Suppose that $\Delta>0$. Under Assumptions (A2)-(A6) and (B1$^{\prime}$)-(B2$^{\prime}$), the  rejection rule in (\ref{multichange-point-rule}) satisfies that \begin{align*}
\lim_{n\to\infty}\P\left(S_n^{\dagger}>\Delta+Q_{1-\alpha,\epsilon}U_{n,\epsilon}^{\dagger}\right)= \begin{cases}
0  & \text{ if }  \sum_{k=1}^{K}\int \delta_k^2(x)dx< \Delta, \\
\alpha  & \text{ if } \sum_{k=1}^{K}\int \delta_k^2(x)dx = \Delta \text{ and }   (\star\star),  \\
1  & \text{ if } \sum_{k=1}^{K}\int \delta_k^2(x)dx> \Delta.
\end{cases} 
\end{align*}
\end{theorem}

\section{Numerical experiments}\label{Sec:simulation}
In this section, numerical experiments are conducted to evaluate the finite-sample performance of the proposed testing procedures for one-sample relevant hypothesis, two-sample relevant hypothesis, and relevant change point.  

For the data generating process, we adopt the Karhunen-Lo\`eve representation and model (\ref{model1}) can be rewritten as 
\begin{align}\label{model2}
    Y_{ij} = m(X_{ij}) + \sum_{k=1}^{\infty} \sqrt{\lambda_k}\,\xi_{ik}\,\psi_k(X_{ij}) + \sigma(X_{ij}) \varepsilon_{ij}, 
    \quad 1 \leq i \leq n,\; 1 \leq j \leq N_i,
\end{align}
where $\lambda_{1} \geq \lambda_{2} \geq \cdots \geq 0$ with $\sum_{k=1}^\infty \lambda_k < \infty$, and $\{\psi_k\}_{k=1}^\infty$ form an orthonormal basis of $\mathcal{L}^2[0,1]$. The FPC scores $\{\xi_{ik}\}_{k=1}^\infty$ are uncorrelated with mean zero and unit variance. In our general setting, the eigenvalues are $\lambda_k = 2^{1-k}$ for $k \leq 4$ and $\lambda_k = 0$ for $k \geq 5$, with eigenfunctions $\psi_{2k-1}(x) = \sqrt{2}\sin(2k\pi x)$ and $\psi_{2k}(x) = \sqrt{2}\cos(2k\pi x)$ for $k \in \mathbb{Z}_+$. The FPC scores $\{\xi_{tk}\}_{t=1,k=1}^{n,4}$ follow an MA(1) process
$\xi_{tk} = 0.8\,\zeta_{tk} + 0.6\,\zeta_{t-1,k}$,
where $\{\zeta_{tk}\}$ are i.i.d. random variables. The design points $X_{ij} \stackrel{\text{i.i.d.}}{\sim} U(0,1)$, and $\varepsilon_{ij} \stackrel{\text{i.i.d.}}{\sim} N(0,1)$ with $\sigma^2(x) \equiv 1$.  Three distributions for the innovations $\zeta_{tk}$ are considered:  
the standard normal distribution $N(0, 1)$, the standardized uniform distribution $U(-\sqrt{3}, \sqrt{3})$, and the standardized Laplace distribution with density $f(x) = 2^{-1/2} \exp(-\sqrt{2}|x|)$. 
To assess the method under varying sampling frequencies,  we consider four observation schemes from sparse to dense, with $N_i$ drawn independently from a discrete uniform distribution over:  
(1) $\{3,4,5,6\}$;  
(2) $\{\lfloor 2n^{1/5} \rfloor, \ldots, \lfloor 4n^{1/5} \rfloor\}$;  
(3) $\{\lfloor n^{1/2} \rfloor, \ldots, \lfloor 2n^{1/2} \rfloor\}$;  
(4) $\{\lfloor n/8 \rfloor, \ldots, \lfloor n/4 \rfloor\}$.

\par For  one-sample problems, we consider mean function as
$$
m(x)= \sqrt a \left(2\sqrt{3}\pi  /\sqrt{2\pi^2-3} \right)x\sin(\pi x),
$$ 
which ensures $\left\|m\right\|_{\mathcal{L}^2}^2  = a$. 
In two-sample problems, we consider $n_1=n$, $n_2=1.2n$, and the mean functions as 
$$
    m_1(x)=  m_2(x)  + \sqrt{3a}x = \left(2\sqrt{3}\pi  /\sqrt{2\pi^2-3} \right)x\sin(\pi x) + \sqrt{3a}x,
$$ 
which guarantees $ \left\|m_1 - m_2\right\|_{\mathcal{L}^2}^2 = a$. 
  For relevant change point problems, we set the location of the change point as $k_n/n=0.4$, and evaluate 
two types of the jump function: (i) $\Delta_{a}(x)\equiv \sqrt a$, (ii) $\Delta_a(x)\equiv 4\sqrt{5a}(x-1/2)^2$,  
each satisfying that $\left\|\Delta_a\right\|_{\mathcal{L}^2}^2=a$.  
The sample sizes $n$ are  $100$, $200$,  $400$, the default trimming parameter $\epsilon$ is $0.1$, and the number of  Monte Carlo replications is  $500$.

\par In our simulations, the cubic splines with $p=4$ are adopted.  The number of knots $J_n$ is a crucial smoothing
 parameter and a data-driven method based on the BIC criterion is what we recommend. For the one-sample problem as an example, we select the integer  $J_n$ in accordance with Assumption (A6)    in the range  
 $$
 J_n \in [ 0.5n^{1/8}(\log n)^{1/4}, \min\{  n^{1/8}\bar{N}^{7/8},2n^{3/10}\}],
 $$
 and the BIC value is defined by
\begin{align*} 
\mathrm{BIC}\left(J_{n}\right) &= \log\left(\frac{1}{n}\sum_{i = 1}^{n}\frac{1}{N_i}\sum_{j = 1}^{N_i}\left[\left\{Y_{ij} - \widehat{m}_{J_n}(X_{ij})\right\}^2  \right] \right) + \frac{(J_n+p)\log(n)}{n},
\end{align*}
where 
$\bar N=\left(n^{-1}\sum_{i=1}^{n}N_i^{-1}\right)^{-1}$ is the harmonic mean of $N_i$'s, and $\widehat{m}_{J_n}(\cdot)$ is defined in (\ref{DEF:m_hat}),  with the number $J_n$ of interior knots.   For the two-sample and relevant change problems, the BIC value can be modified accordingly without difficulty.

The results are displayed in Figures  \ref{fig:Onesample} and S.1-S.3
in the Supplementary Materials, from which we draw the following conclusions.  First, under the null hypothesis, the empirical size of the test is well controlled at the prespecified significance level 
$\alpha=0.05$. Second, the empirical power increases as the sample size grows or the departure from the null becomes larger. Moreover, the proposed method performs robustly  for discretely observed functional data with various distributions and different sampling frequencies.

To evaluate the  sensitivity of $\epsilon$ and alternative self-normalizers, we conduct additional simulations in Sections S.3.1-S.3.2
of the Supplementary Materials.

\begin{figure}
    \centering
    \includegraphics[width=0.31\linewidth]{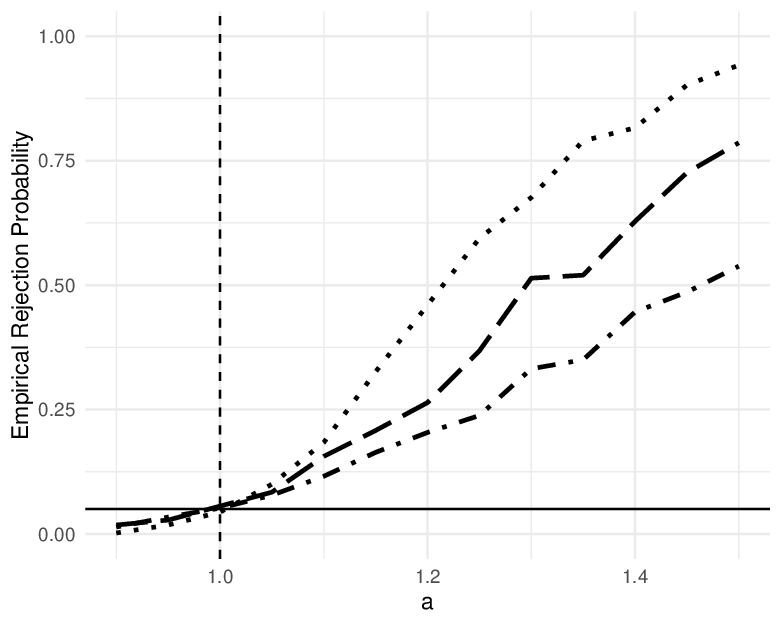}
    \includegraphics[width=0.31\linewidth]{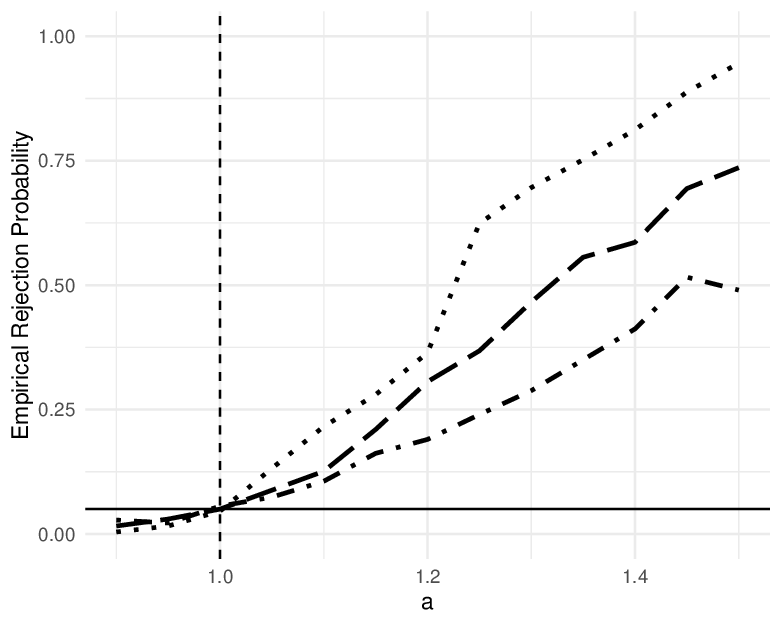}
    \includegraphics[width=0.31\linewidth]{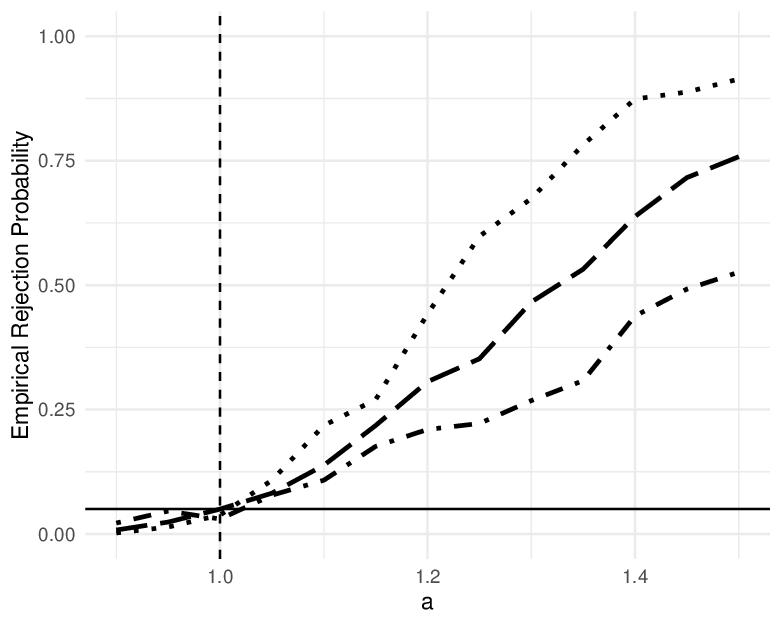}
    \includegraphics[width=0.31\linewidth]{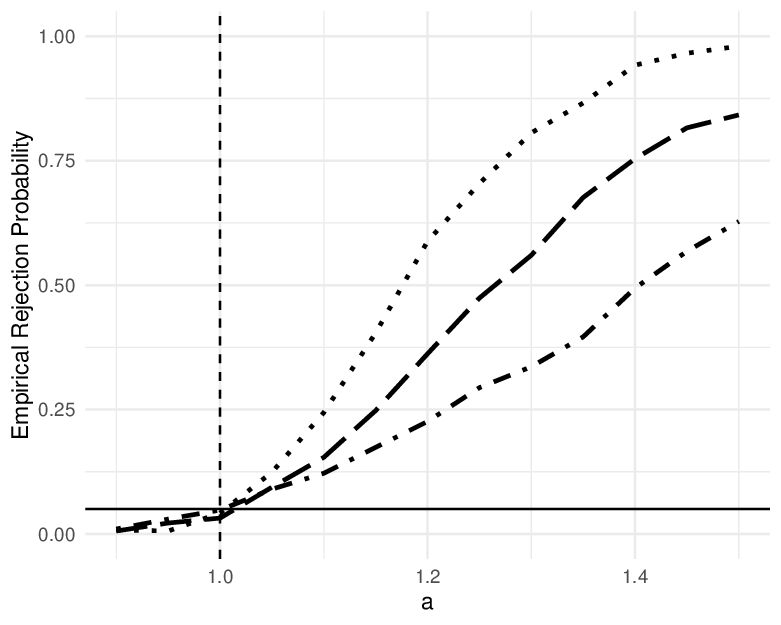}
    \includegraphics[width=0.31\linewidth]{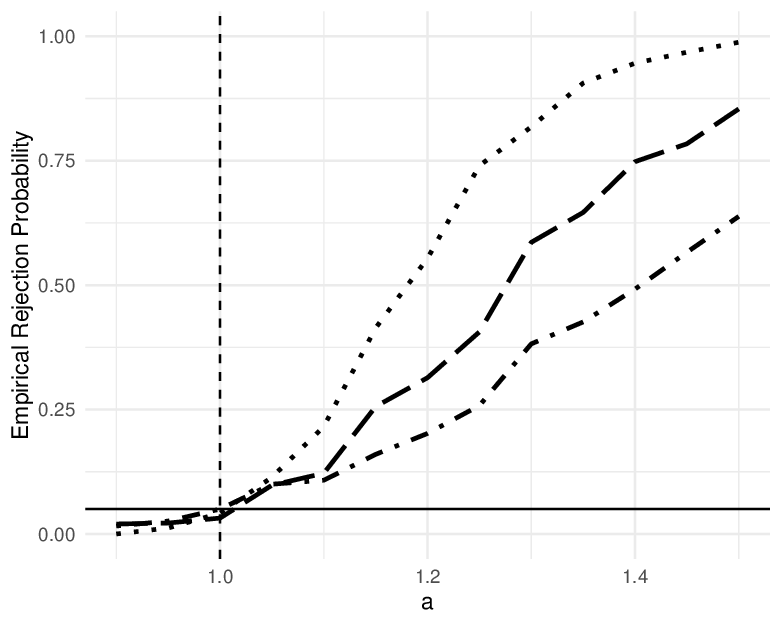}
    \includegraphics[width=0.31\linewidth]{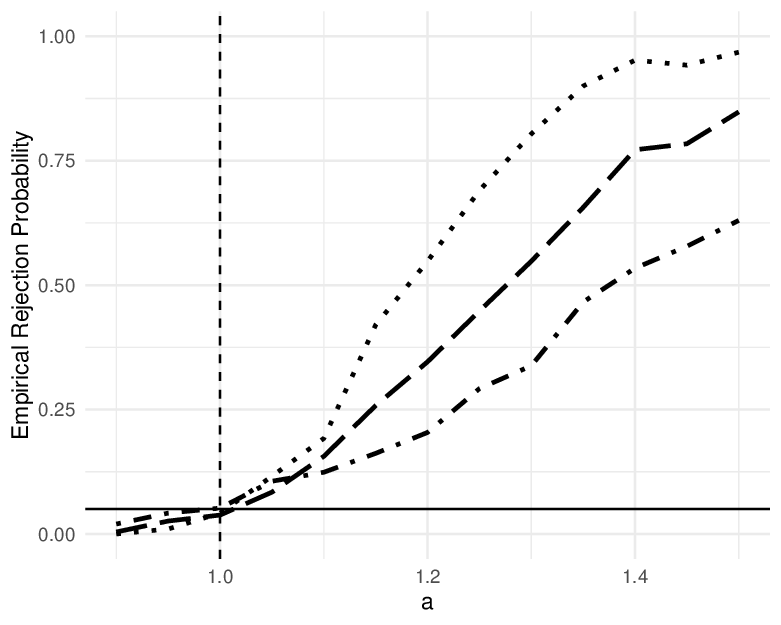}
    \includegraphics[width=0.31\linewidth]{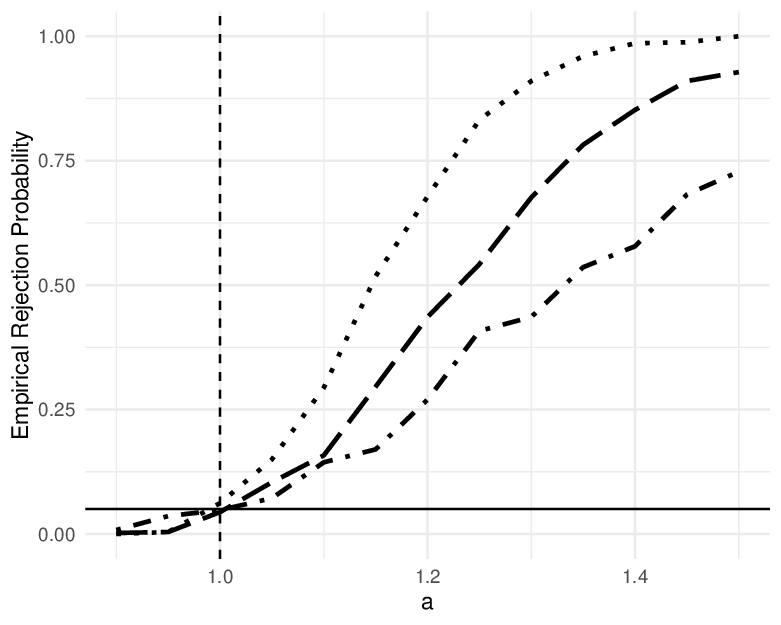}
    \includegraphics[width=0.31\linewidth]{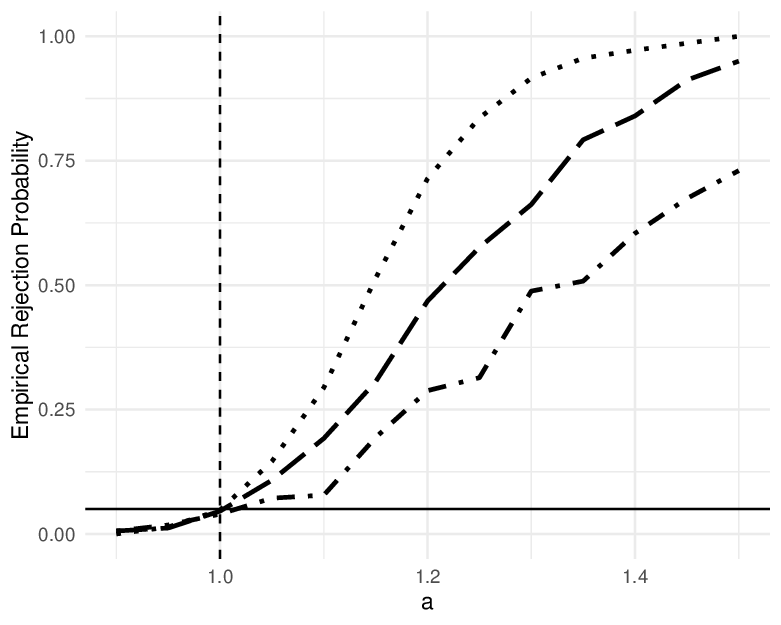}
    \includegraphics[width=0.31\linewidth]{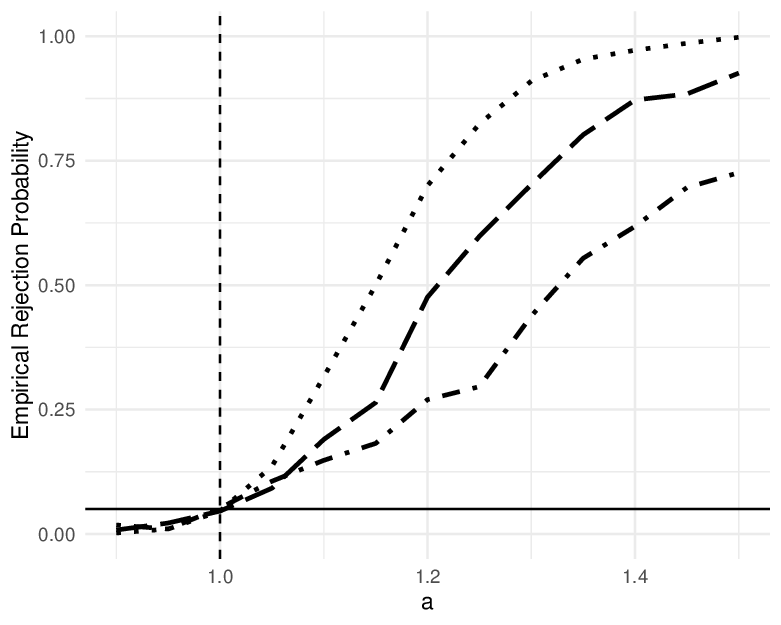}
    \includegraphics[width=0.31\linewidth]{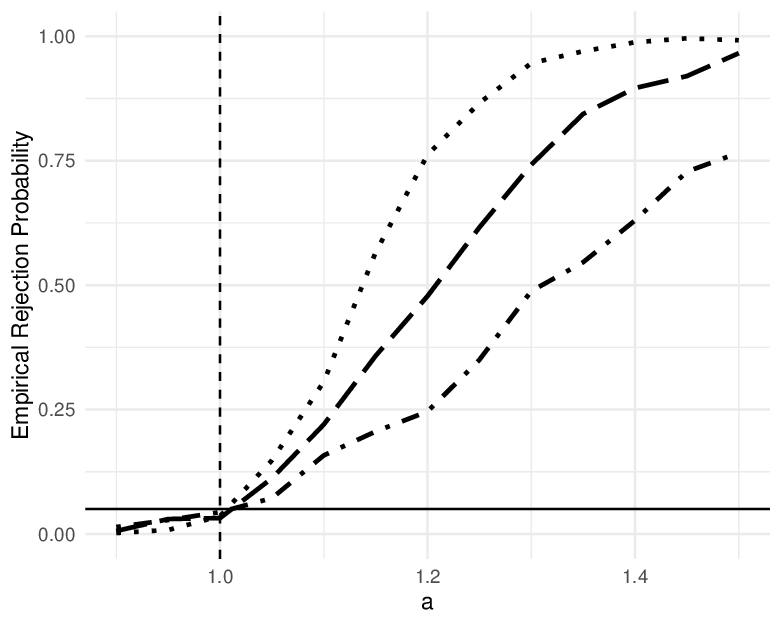}
    \includegraphics[width=0.31\linewidth]{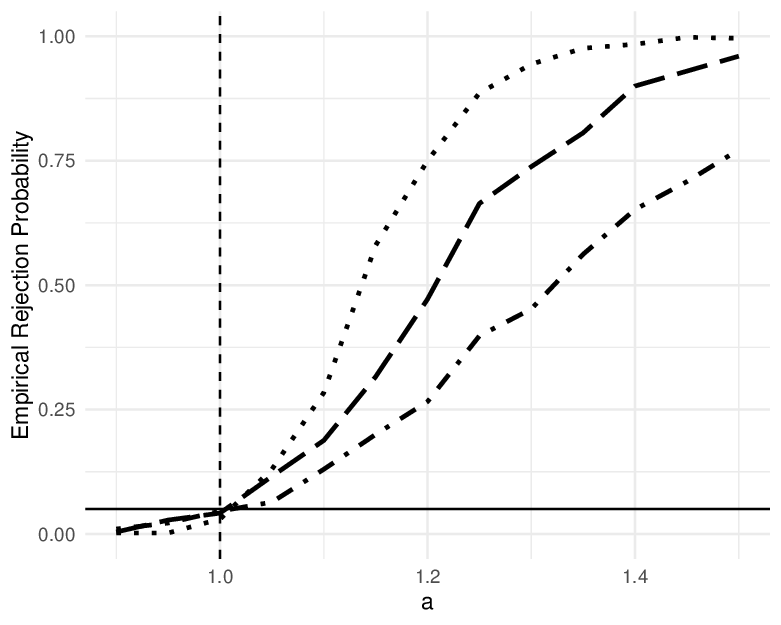}
\includegraphics[width=0.31\linewidth]{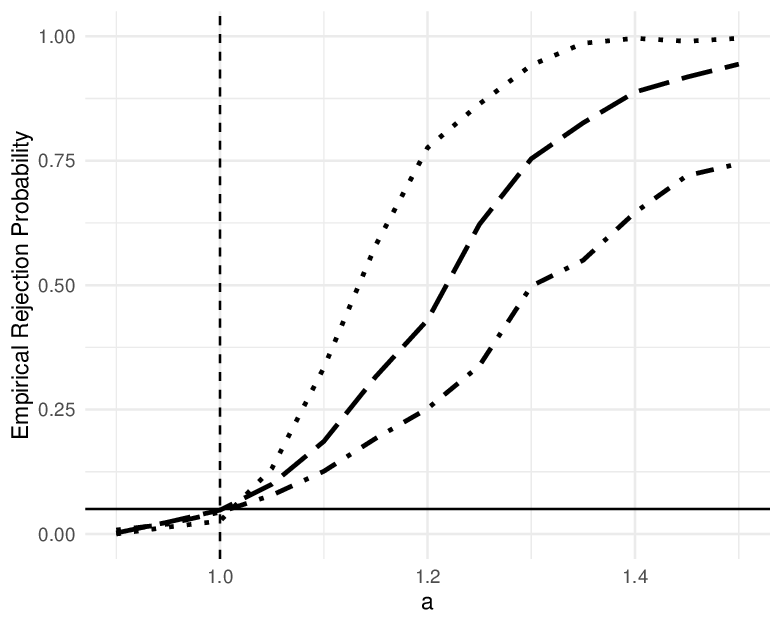}
\caption{Empirical rejection probabilities of the one-sample test   for the relevant hypotheses (\ref{one-sample-test}) with $\Delta=1$. Different sampling schemes (rows) and different distributions of the innovations (columns) are considered. Dotdash lines: $n=100$; longdash lines: $n=200$; dotted lines: $n=400$. The first row: scenario (1); the second row: scenario (2); the third row: scenario (3); the last row:scenario (4). Left: normal; middle: uniform; right: Laplace. }
    \label{fig:Onesample}
\end{figure}

\section{Application}

\subsection{Sparse case: AU.SHF data}
\par In this subsection, implied volatility (IV) curves for options on the AU.SHF index (gold options listed on the Shanghai Futures Exchange) are used to illustrate our methodology. Such contracts confer the right to buy AU.SHF at a predetermined strike price on or before a specified expiration date. Implied volatility, obtained from the Black-Scholes formula from observed option prices, together with the risk-free rate and time to maturity, is commonly modeled as a function of moneyness (strike divided by the underlying price). 
\par IV both provides a diagnostic for the Black-Scholes assumptions and summarizes market sentiment; see \citet{hull2016options}. Following \citet{lam2011estimation}, we work with the first-order differences of the IV curves, producing $n=709$ functional observations from May 16, 2022, to April 16, 2025. Although these curves are often treated as fully observed functional time series, they are constructed by interpolation and therefore subject to measurement error, \citet{liu2016convolutional,dette2021confidence}. For implementation, we discretize each curve on $N_i\equiv 29$ equally spaced grid points over the moneyness interval $[0.6,1.3]$, which is then further scaled to $[0,1]$.

\par Applying the testing and detection procedure in \citet{cai2024detecting}, one obtains the $p$-value  $<0.001$, implying there exists a change point in IV curves, and $\widehat k_n = 404$ with date January 5, 2024.  On January 5, 2024, as trading desks squared positions for quarter- and year-end and in anticipation of the New Year’s Day market closure, liquidity in the gold‐options market dried up. Dealers demanded a higher risk premium to carry gamma exposure over the holiday, driving implied volatilities sharply higher on the final trading day of the year. Seasonal anomalies of this kind in China’s gold market have been documented by \citet{qi2013monthly}, who report abnormally elevated volatility and returns around December in Shanghai gold contracts.

\par Therefore, we are interested in the relevant change in New Year’s Day market. Table \ref{tab:realdata1} shows that there is no strong evidence that the integrated squared mean difference exceeds $3.4$, whereas there is clear evidence that it exceeds $2.0$. Choosing $\Delta$ in the range $[2.1,2.9]$ allows us to reject the null hypothesis at the $5\%$  significance level, while choosing $\Delta$ in the range $[3.0,3.3]$ allows us to reject it at the $10\%$ level.

\begin{figure}
    \centering
    \includegraphics[width=0.48\linewidth]{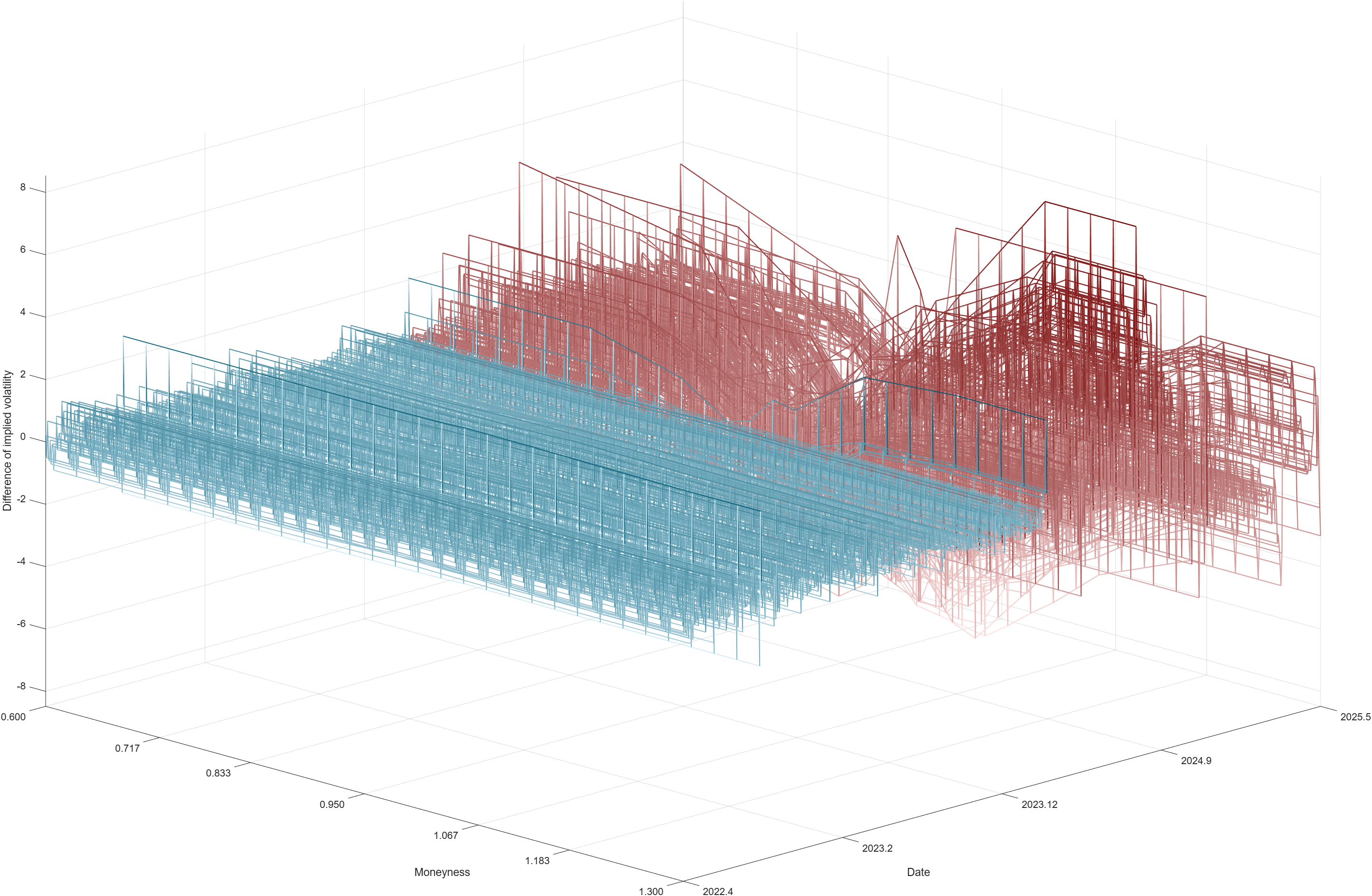}
    \includegraphics[width=0.48\linewidth]{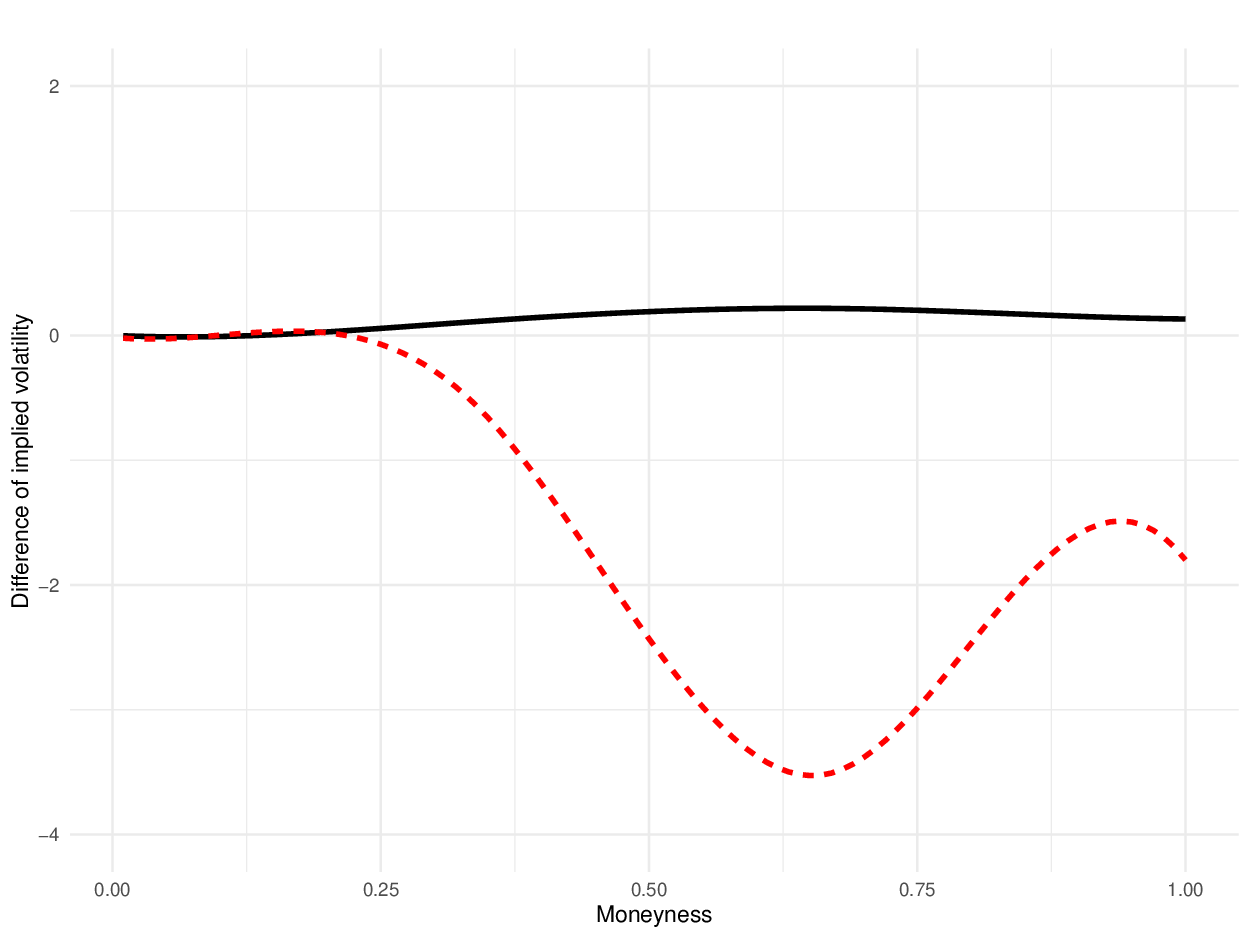}
    \caption{Left panel: the first order difference of AU.SHF implied volatility before and after the change point January 5, 2024. Right panel: the functional mean estimation of the two corresponding  periods (line: before, dotted: after). }
    \label{fig:realdata1}
\end{figure}

\begin{table}[htbp]
  \centering
  \caption{Summary of the  tests for relevant hypotheses in (\ref{relevantchange-test}) with varying $\Delta$ for the financial data. }
    \begin{tabular}{cccc}
    \toprule
    $\Delta$ & $90\%$  quantile & $95\%$  quantile & $99\%$  quantile \\
    \midrule
    3.4   & False & False & False \\
    3.3   & True & False & False \\
    3.0   & True  & False & False \\
    2.9   & True  & True  & False \\
    2.1   & True  & True  & False \\
    2.0  & True  & True  & True \\
    \bottomrule
    \end{tabular}%
  \label{tab:realdata1}%
\end{table}%

\subsection{Dense case: traffic volume data}

The second dataset is publicly available from the University of California, Irvine Machine Learning Repository \url{https://archive.ics.uci.edu/dataset/157/dodgers+loop+sensor} and has been analyzed in 
\citet{zhang2015varying} and \citet{cai2024sparse}. It was collected by a loop sensor installed on the Glendale on ramp of the $101$ North freeway in Los Angeles, near Dodger Stadium, which is the home field of the Los Angeles Dodgers baseball team. The sensor continuously monitored traffic volume, recording the total number of vehicles every five minutes. Our analysis focuses on whether there is a significant difference in traffic patterns  between days with a Dodgers home game and those without.

The data span from April to September 2005. After handling missing values, the observations are divided into two groups based on whether a Dodgers home game took place on a given day. The first sample  consists of traffic flow data on days with a home game ($n_1 = 75$), while the second sample  includes traffic flow data on days without a home game ($n_2 = 69$). For both samples, each day contains $N_i = 288$ measurements, corresponding to five-minute intervals over 24 hours. 

\begin{figure}
    \centering
    \includegraphics[width=0.48\linewidth]{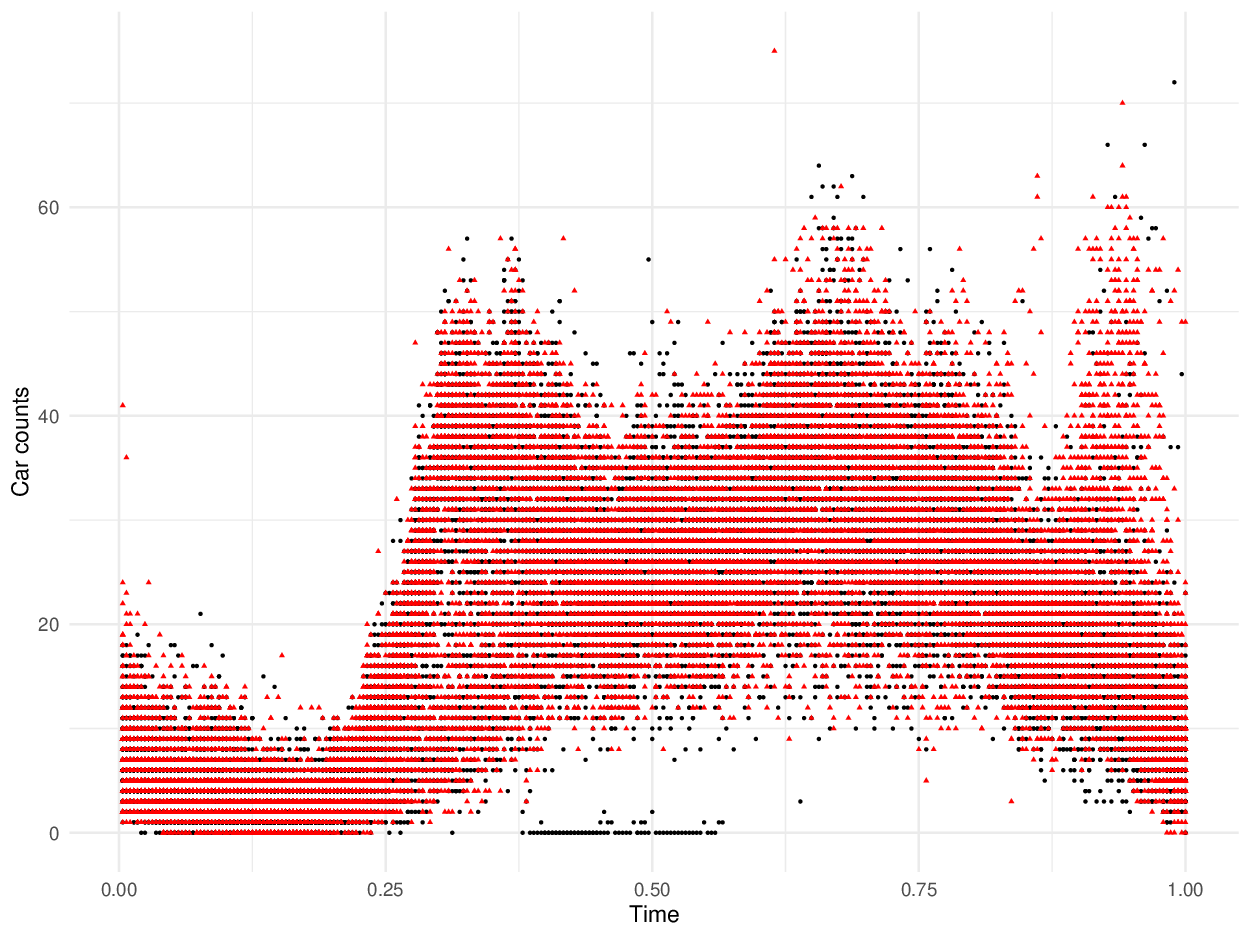}
    \includegraphics[width=0.48\linewidth]{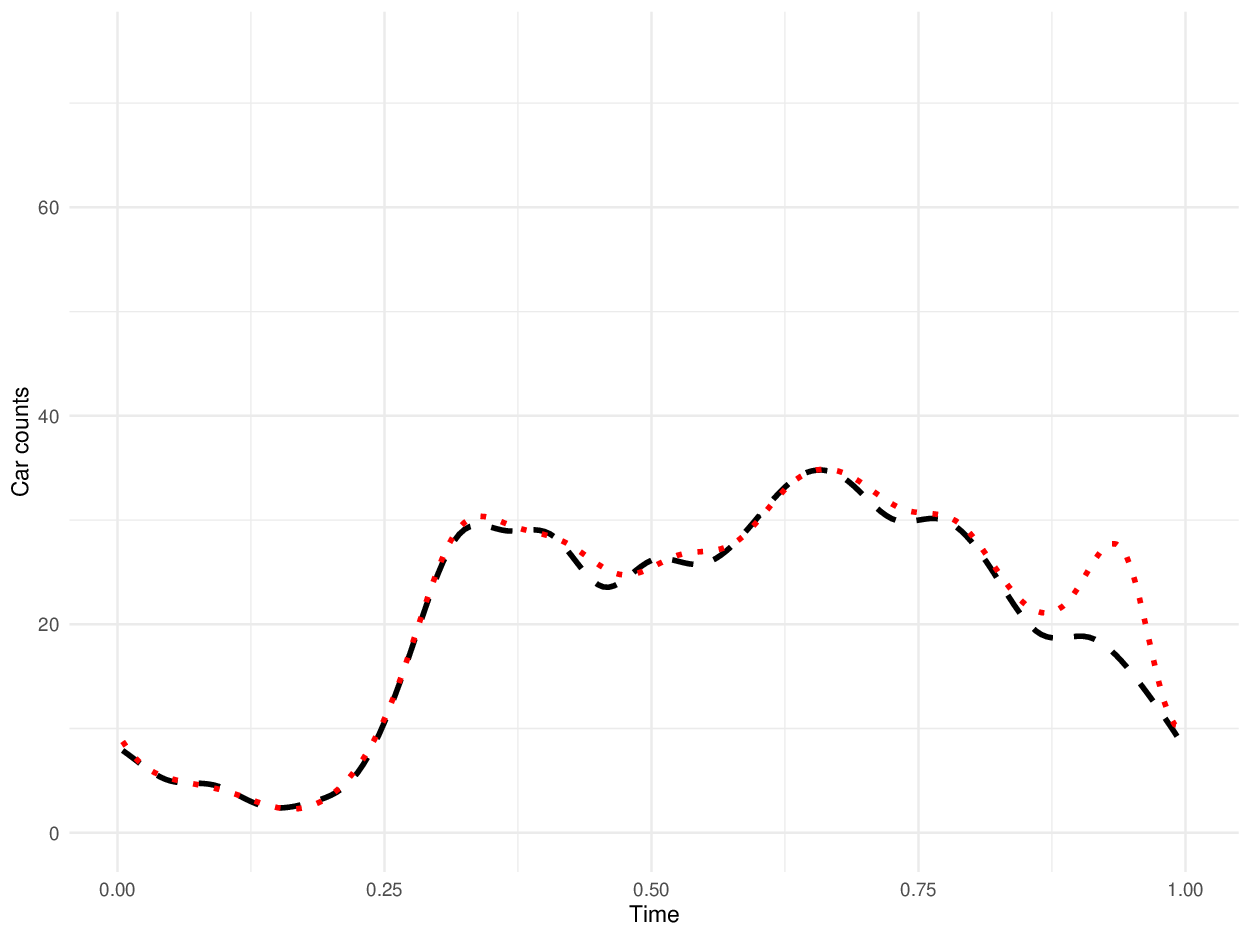}
    \caption{Left panel: car counts recorded every 5 minutes over 144 days; circles:  days without home games,  triangles:  days with home games.  
Right panel: the estimated mean function using cubic B-splines; dashed line:  days without home games,  dotted line:  days with home games. 
}
    \label{fig:realdata2}
\end{figure}

Since baseball games are typically held in the evening,  traffic on days with a home game tends to be heavier during the evening hours compared to days without a home game, as shown in Figure \ref{fig:realdata2}. To further assess the magnitude of this difference, we perform the two-sample relevant test in (\ref{two-sample-test}) and report the test decisions for several choices of the significance level $\alpha$ and the threshold parameter $\Delta$ in Table \ref{tab:realdata2}. The results indicate no strong evidence of an integrated squared mean difference exceeding $3.8$. However, one can assert with at least $95\%$  confidence that the integrated squared mean difference exceeds $3.0$.

\begin{table}[h!]
  \centering
  \caption{Summary of the  tests for relevant hypotheses in (\ref{two-sample-test}) with varying $\Delta$ for the traffic data. }
    \begin{tabular}{cccc}
    \toprule
    $\Delta$ & $90\%$  quantile & $95\%$  quantile & $99\%$  quantile \\
    \midrule
    3.8   & False & False & False \\
    3.7   & True  & False & False \\
    3.1  & True  & False & False \\
    3.0   & True  & True  & False \\
    1.4   & True  & True  & False \\
    1.3   & True  & True  & False \\
    \bottomrule
    \end{tabular}%
  \label{tab:realdata2}%
\end{table}%

\section{Concluding remark}
\par In this paper, we developed the unified theory of relevant hypothesis tests for functional time series, which address the one sample, two sample, relevant single change point and multiple change point problems. The key features are our methodologies are built from discretely observed trajectories with measurement error, and enjoy the flexibility for arbitrary sampling strategies with unified test statistics and asymptotic distribution. The theoretical results establish the consistency of the proposed test statistics, including the size and power, and derive the sparse-to-dense phase transition through the asymptotic variance of the scalar target estimators. Further, we discuss  the selection of self-normalizers and consider a power-enhanced candidate with range adjustment. Extensive simulations provide direct evidence of correctness and robustness of developed methodologies. The implied volatility and traffic volume data illustrations present some interesting findings. 
Despite these contributions, there are also some limitations, including  stationarity and low dimensionality. The local stationary or non-stationary functional time series with higher dimension will be more applicable in real world, while the nonstationarity and additional diverging factor bring significant challenge in self-normalization  and Gaussian approximation, which we remain as future directions.

\bibliographystyle{abbrvnat}
\bibliography{ref}

@article{zhou2018efficient,
  title={Efficient estimation of the nonparametric mean and covariance functions for longitudinal and sparse functional data},
  author={Zhou, Ling and Lin, Huazhen and Liang, Hua},
  journal={Journal of the American Statistical Association},
  volume={113},
  number={524},
  pages={1550--1564},
  year={2018},
  publisher={Taylor \& Francis}
}

@article{6dd3c0f4-00c9-3011-90a1-202033e64598,
 ISSN = {01621459, 1537274X},
 author = {Alexander M. Samarov},
 journal = {Journal of the American Statistical Association},
 number = {423},
 pages = {836--847},
 publisher = {[American Statistical Association, Taylor & Francis, Ltd.]},
 title = {Exploring Regression Structure Using Nonparametric Functional Estimation},
 urldate = {2026-05-27},
 volume = {88},
 year = {1993}
}

@article{bickel1988estimating,
  title={Estimating integrated squared density derivatives: sharp best order of convergence estimates},
  author={Bickel, Peter J and Ritov, Yaacov},
  journal={Sankhy{\=a}: The Indian Journal of Statistics, Series A},
  pages={381--393},
  year={1988},
  publisher={JSTOR}
}

@article{zhou2023statistical,
  title={Statistical inference for high-dimensional panel functional time series},
  author={Zhou, Zhou and Dette, Holger},
  journal={Journal of the Royal Statistical Society Series B: Statistical Methodology},
  volume={85},
  number={2},
  pages={523-549},
  year={2023},
  publisher={Oxford University Press US}
}

@article{guo2025sparse,
  title={From sparse to dense functional data in high dimensions: Revisiting phase transitions from a non-asymptotic perspective},
  author={Guo, Shaojun and Li, Dong and Qiao, Xinghao and Wang, Yizhu},
  journal={Journal of Machine Learning Research},
  volume={26},
  number={15},
  pages={1--40},
  year={2025}
}

@article{cai2024detecting,
  title={From sparse to dense functional time series: phase transitions of detecting structural breaks and beyond},
  author={Cai, Leheng and Hu, Qirui},
  journal={arXiv preprint arXiv:2412.20858},
  year={2024}
}

@article{cao2012simultaneous,
  title={Simultaneous inference for the mean function based on dense functional data},
  author={Cao, Guanqun and Yang, Lijian and Todem, David},
  journal={Journal of nonparametric statistics},
  volume={24},
  number={2},
  pages={359--377},
  year={2012},
  publisher={Taylor \& Francis}
}

@book{bosq2000linear,
  title={Linear processes in function spaces: theory and applications},
  author={Bosq, Denis},
  volume={149},
  year={2000},
  publisher={Springer Science \& Business Media}
}

@article{2022lijie,
  title={Statistical inference for functional time series},
  author={Li, Jie and Yang, Lijian},
  journal={Statistica Sinica},
  volume={33},
  number={1},
  pages={519--549},
  year={2023},
  publisher={JSTOR}
}

@article{Zhang2021UnifiedPC,
  title={Unified principal component analysis for sparse and dense functional data under spatial dependency},
  author={Zhang, Haozhe and Li, Yehua},
  journal={Journal of Business \& Economic Statistics},
  volume={40},
  number={4},
  pages={1523--1537},
  year={2022},
  publisher={Taylor \& Francis}
}

@article{yao2005functional,
  title={Functional data analysis for sparse longitudinal data},
  author={Yao, Fang and M{\"u}ller, Hans-Georg and Wang, Jane-Ling},
  journal={Journal of the American statistical association},
  volume={100},
  number={470},
  pages={577--590},
  year={2005},
  publisher={Taylor \& Francis}
}

@Article{ZW16,
  author  = {Zhang, X. and Wang, J.-L.},
  title   = {From sparse to dense functional data and beyond},
  journal = {The Annals of Statistics},
  year    = {2016},
  volume  = {44},
  pages   = {2281-2321},
}

@article{dette2020testing,
  title={Testing relevant hypotheses in functional time series via self-normalization},
  author={Dette, Holger and Kokot, Kevin and Volgushev, Stanislav},
  journal={Journal of the Royal Statistical Society Series B: Statistical Methodology},
  volume={82},
  number={3},
  pages={629--660},
  year={2020},
  publisher={Oxford University Press}
}

@article{zheng2014smooth,
  title={A smooth simultaneous confidence corridor for the mean of sparse functional data},
  author={Zheng, Shuzhuan and Yang, Lijian and H{\"a}rdle, Wolfgang K},
  journal={Journal of the American Statistical Association},
  volume={109},
  number={506},
  pages={661--673},
  year={2014},
  publisher={Taylor \& Francis}
}

@article{tonycai2011,
author = {T. Tony Cai and Ming Yuan},
title = {{Optimal estimation of the mean function based on discretely sampled functional data: Phase transition}},
volume = {39},
journal = {The Annals of Statistics},
number = {5},
publisher = {Institute of Mathematical Statistics},
pages = {2330 -- 2355},
year = {2011}
}

@article{zhang2015varying,
  title={Varying-coefficient additive models for functional data},
  author={Zhang, Xiaoke and Wang, Jane-Ling},
  journal={Biometrika},
  volume={102},
  number={1},
  pages={15--32},
  year={2015},
  publisher={Oxford University Press}
}

@article{sharghi2021mean,
  title={On mean derivative estimation of longitudinal and functional data: from sparse to dense},
  author={Sharghi Ghale-Joogh, Hassan and Hosseini-Nasab, S Mohammad E},
  journal={Statistical Papers},
  volume={62},
  number={4},
  pages={2047--2066},
  year={2021},
  publisher={Springer}
}

@article{cai2024sparse,
  title={From sparse to dense functional data: Phase transitions from a simultaneous inference perspective},
  author={Cai, Leheng and Hu, Qirui},
  journal={Statistics and Computing},
  volume={35},
  number={5},
  pages={1--19},
  year={2025},
  publisher={Springer}
}

@book{horvath2012inference,
  title={Inference for functional data with applications},
  author={Horv{\'a}th, Lajos and Kokoszka, Piotr},
  volume={200},
  year={2012},
  publisher={Springer Science \& Business Media}
}

@article{madrid2022change,
  title={Change-point detection for sparse and dense functional data in general dimensions},
  author={Madrid Padilla, Carlos Misael and Wang, Daren and Zhao, Zifeng and Yu, Yi},
  journal={Advances in Neural Information Processing Systems},
  volume={35},
  pages={37121--37133},
  year={2022}
}

@article{cai2024global,
   title={Global inference and test for eigensystems of imaging data over complicated domains},
  author={Cai, Leheng and Hu, Qirui},
  journal={Journal of Computational and Graphical Statistics},
  volume={34},
  number={2},
  pages={729--745},
  year={2025},
  publisher={Taylor \& Francis}
}

@article{dette2016detecting,
  title={Detecting relevant changes in time series models},
  author={Dette, Holger and Wied, Dominik},
  journal={Journal of the Royal Statistical Society Series B: Statistical Methodology},
  volume={78},
  number={2},
  pages={371--394},
  year={2016},
  publisher={Oxford University Press}
}

@article{DK21,
  author = {Dette, H. and Kutta, T.},
  title = {Detecting structural breaks in eigensystems of functional time series},
  journal = {Electron. J. Stat.},
  year = {2021},
  volume = {15},
  number = {1},
  pages = {944-983}
}

@article{dette2020functional,
  title={Functional data analysis in the Banach space of continuous functions},
  author={Dette, Holger and Kokot, Kevin and Aue, Alexander},
  journal={The Annals of Statistics},
  volume={48},
  number={2},
  pages={1168--1192},
  year={2020},
  publisher={JSTOR}
}

@article{van2024general,
  title={A general framework to quantify deviations from structural assumptions in the analysis of nonstationary function-valued processes},
  author={Van Delft, Anne and Dette, Holger},
  journal={The Annals of Statistics},
  volume={52},
  number={2},
  pages={550--579},
  year={2024},
  publisher={Institute of Mathematical Statistics}
}

@article{bucher2021deviations,
  title={Are deviations in a gradually varying mean relevant? A testing approach based on sup-norm estimators},
  author={B{\"u}cher, Axel and Dette, Holger and Heinrichs, Florian},
  journal={The Annals of Statistics},
  volume={49},
  number={6},
  pages={3583--3617},
  year={2021},
  publisher={Institute of Mathematical Statistics}
}

@article{dette2021bio,
  title={Bio-equivalence tests in functional data by maximum deviation},
  author={Dette, Holger and Kokot, Kevin},
  journal={Biometrika},
  volume={108},
  number={4},
  pages={895--913},
  year={2021},
  publisher={Oxford University Press}
}

@article{van2022pivotal,
  title={Pivotal tests for relevant differences in the second order dynamics of functional time series},
  author={Van Delft, Anne and Dette, Holger},
  journal={Bernoulli},
  volume={28},
  number={4},
  pages={2260--2293},
  year={2022},
  publisher={Bernoulli Society for Mathematical Statistics and Probability}
}

@article{LWSTD2025,
  title={Simultaneous Inference for Mean Curves of Functional
 and Longitudinal Data: A Unified Theory},
  author={Zhipeng Luo and Weibiao Wu},
  journal={Statistica Sinica},
  year={2025+}
}

@article{kim2013unified,
  title={Unified inference for sparse and dense longitudinal models},
  author={Kim, Seonjin and Zhao, Zhibiao},
  journal={Biometrika},
  volume={100},
  number={1},
  pages={203--212},
  year={2013},
  publisher={Oxford University Press}
}

@article{BD87,
  author = {Berger, J. O. and Delampady, M.},
  title = {Testing precise hypotheses},
  journal = {Statistical Science},
  year = {1987},
  pages = {317-335}
}

@article{B38,
  author = {Berkson, J.},
  title = {Some difficulties of interpretation encountered in the application of the chi-square test},
  journal = {J. Am. Statist. Ass},
  year = {1938},
  volume = {33},
  number = {203},
  pages = {526-536}
}

@book{hull2016options,
  title={Options, futures, and other derivatives},
  author={Hull, John C and Basu, Sankarshan},
  year={2016},
  publisher={Pearson Education India}
}

@article{qi2013monthly,
  title={The monthly effects in Chinese gold market},
  author={Qi, Ming and Wang, Wenyao},
  journal={International Journal of Economics and Finance},
  volume={5},
  number={10},
  pages={141--146},
  year={2013}
}

@article{lam2011estimation,
  title={Estimation of latent factors for high-dimensional time series},
  author={Lam, Clifford and Yao, Qiwei and Bathia, Neil},
  journal={Biometrika},
  volume={98},
  number={4},
  pages={901--918},
  year={2011},
  publisher={Oxford University Press}
}

@article{liu2016convolutional,
  title={Convolutional autoregressive models for functional time series},
  author={Liu, Xialu and Xiao, Han and Chen, Rong},
  journal={Journal of Econometrics},
  volume={194},
  number={2},
  pages={263--282},
  year={2016},
  publisher={Elsevier}
}

@article{dette2021confidence,
  author={Dette, Holger and Wu, Weichi},
  title={Confidence surfaces for the mean of locally stationary functional time series},
  journal={Statistica Sinica},
  year={2026},
    volume={36},
  pages={583-604}
}

@article{shao2015self,
  title={Self-normalization for time series: a review of recent developments},
  author={Shao, Xiaofeng},
  journal={Journal of the American Statistical Association},
  volume={110},
  number={512},
  pages={1797--1817},
  year={2015},
  publisher={Taylor \& Francis}
}

@article{shao2010self,
  title={A self-normalized approach to confidence interval construction in time series},
  author={Shao, Xiaofeng},
  journal={Journal of the Royal Statistical Society: Series B (Statistical Methodology)},
  volume={72},
  number={3},
  pages={343--366},
  year={2010},
  publisher={Wiley Online Library}
}

@article{berger2025dense,
  title={From dense to sparse design: Optimal rates under the supremum norm for estimating the mean function in functional data analysis},
  author={Berger, Max and Hermann, Philipp and Holzmann, Hajo},
  journal={Bernoulli},
  volume={31},
  number={3},
  pages={1858--1888},
  year={2025},
  publisher={Bernoulli Society for Mathematical Statistics and Probability}
}

\end{document}